\documentclass[journal]{IEEEtran}

\usepackage{lineno,hyperref}
\usepackage{times}
\usepackage{epsfig}
\usepackage{graphicx}
\usepackage{amsmath}
\usepackage{amssymb}
\usepackage{subfigure}
\usepackage{multirow}
\usepackage{multicol}
\usepackage{booktabs}
\usepackage{enumerate}
\usepackage{algorithm}
\usepackage{algpseudocode}
\usepackage{amsmath}
\usepackage{float}
\usepackage{array}
\usepackage{cite}
\usepackage{color}
\usepackage{bookmark}
\usepackage{changepage}

\modulolinenumbers[5]

\begin{document}

\title{SSCU-Net: Spatial-Spectral Collaborative Unmixing Network for Hyperspectral Images}

\author{Lin~Qi,
        Feng~Gao,~\IEEEmembership{Member,~IEEE},
        Junyu~Dong,~\IEEEmembership{Member,~IEEE},
        Xinbo~Gao,~\IEEEmembership{Senior Member,~IEEE},
        and~
        Qian~Du,~\IEEEmembership{Fellow,~IEEE}
\thanks{This work was supported in part by the National Key Research and Development Program of China under Grant 2018AAA0100602, in part by the National Natural Science Foundation of China under Grant 42106191, and in part by the China Postdoctoral Science Foundation under Grant 2021M693023. (Corresponding author: Feng Gao; Junyu Dong.)}
\thanks{Lin Qi, Feng Gao, and Junyu Dong are with the School of Computer Science and Technology, Ocean University of China, Qingdao 266100, China (e-mail: qilin2020@ouc.edu.cn; gaofeng@ouc.edu.cn; dongjunyu@ouc.edu.cn).}
\thanks{Xinbo Gao is with the Chongqing Key Laboratory of Image Cognition, Chongqing University of Posts and Telecommunications, Chongqing 400065, China (e-mail: gaoxb@cqupt.edu.cn).}
\thanks{Qian Du is with the Department of Electrical and Computer Engineering, Mississippi State University, Starkville, MS 39762 USA (e-mail: du@ece.msstate.edu).}
}

\maketitle

\begin{abstract}
Linear spectral unmixing is an essential technique in hyperspectral image processing and interpretation. In recent years, deep learning-based approaches have shown great promise in hyperspectral unmixing, in particular, unsupervised unmixing methods based on autoencoder networks are a recent trend. The autoencoder model, which automatically learns low-dimensional representations (abundances) and reconstructs data with their corresponding bases (endmembers), has achieved superior performance in hyperspectral unmixing. In this article, we explore the effective utilization of spatial and spectral information in autoencoder-based unmixing networks. Important findings on the use of spatial and spectral information in the autoencoder framework are discussed. Inspired by these findings, we propose a spatial-spectral collaborative unmixing network, called SSCU-Net, which learns a spatial autoencoder network and a spectral autoencoder network in an end-to-end manner to more effectively improve the unmixing performance. SSCU-Net is a two-stream deep network and shares an alternating architecture, where the two autoencoder networks are efficiently trained in a collaborative way for estimation of endmembers and abundances. Meanwhile, we propose a new spatial autoencoder network by introducing a superpixel segmentation method based on abundance information, which greatly facilitates the employment of spatial information and improves the accuracy of unmixing network. Moreover, extensive ablation studies are carried out to investigate the performance gain of SSCU-Net. Experimental results on both synthetic and real hyperspectral data sets illustrate the effectiveness and competitiveness of the proposed SSCU-Net compared with several state-of-the-art hyperspectral unmixing methods.
\end{abstract}


\IEEEpeerreviewmaketitle

\section{Introduction}
\IEEEPARstart{D}{ue} to the limited spatial resolution of hyperspectral sensors, together with complex natural surfaces and scattering of the light, mixed pixels, which tend to contain more than one spectral signature, become a challenging issue in hyperspectral image (HSI) processing and application \cite{shippert2004use,plaza2011foreword}. Therefore, hyperspectral unmixing (HU) has been a major technique for the utilization of hyperspectral data, whose purpose is to extract basic features (endmembers) and calculate the corresponding fractions (abundances) in the spectral pixels of HSI \cite{keshava2002spectral}. Relying on different mixing models, unmixing algorithms can be divided into two categories: linear and nonlinear spectral unmixing \cite{keshava2002spectral,bioucas2012hyperspectral}. Linear mixing model (LMM) is widely used in HU due to its computational tractability, and it assumes that the observed spectrum of a mixed pixel is a linear combination of endmember spectra weighted by the corresponding mixing coefficients (abundances) \cite{bioucas2012hyperspectral,ma2014signal}. Subject to physical constraints, abundances need to satisfy two constraints: the abundance nonnegativity constraint (ANC) and the abundance sum-to-one constraint (ASC) \cite{heinz2001fully}.

In this article, we follow most of the previous works and focus on linear spectral unmixing. Under the LMM, early unmixing approaches can be categorized as geometrical, statistical or a sparse regression problem. Depending on whether pure pixels are assumed, the geometrical-based methods can be further divided into pure pixel-based methods or minimum volume-based methods. The best-known representative methods based on pure pixels are the pixel purity index (PPI) \cite{boardman1993automating}, N-FINDR \cite{winter1999n}, and vertex component analysis (VCA) \cite{nascimento2005vertex}. Nevertheless, the assumption of pure pixels in these algorithms is difficult to guarantee in real HSI. Minimum volume-based methods are proposed to process highly mixed hyperspectral data, of which minimum volume simplex analysis (MVSA) \cite{li2015minimum} is a good example. In addition, statistical-based methods are also effective alternatives to process highly mixed HSI and have attracted great attention. For example, the Bayesian framework expresses the unmixing problem as an inference problem, utilizing statistical assumptions and priors to constrain the unmixing results \cite{dobigeon2009joint,eches2010bayesian,nascimento2012hyperspectral}. Due to the unique advantages in learning part-based representation, nonnegative matrix factorization (NMF) is a widely used unmixing tool that can simultaneously estimate endmembers and abundances \cite{qian2011hyperspectral,feng2018hyperspectral,peng2021self}. Many extended NMFs have been proposed, most of which introduce a number of regularization constraints, containing priors into endmember extraction or abundance estimation process to involve spectral and/or spatial information into the NMF framework \cite{zhu2014spectral,wang2017spatial,huang2019spectral}.

Sparse regression-based unmixing approaches are another mainstream. Sparse unmixing \cite{iordache2011sparse,tang2015sparse} is proposed as a semisupervised method, which represents the spectrum of mixed pixel as a sparse linear combination of spectral signatures in a pre-known and large spectral library. The sparse unmixing algorithm via variable splitting and augmented Lagrangian (SUnSAL) \cite{iordache2011sparse} is a typical example of sparse unmixing. Based on the fact that mixed pixels are unlikely to be mixed by a quantity of spectra, it adopts the classic ${l_1}$ regularization term on the abundances. By introducing a total variation (TV) regularization term, the SUnSAL-TV \cite{iordache2012total} algorithm integrates the spatial information between pixels. However, it may cause problems with oversmoothness and blurred boundaries. Thus, collaborative SUnSAL (CLSUnSAL) \cite{iordache2014collaborative} introduces the ${l_{2,1}}$ regularization term and imposes row sparsity on the abundance matrix. On the other hand, in \cite{zheng2016reweighted} and \cite{wang2017hyperspectral}, a weighted sparse unmixing framework with weighted ${l_1}$ norm is proposed, which penalizes the nonzero elements in the abundance matrix by introducing one or more weighting factors. Meanwhile, a variety of spatial and/or spectral weighted regularizers \cite{he2017total,zhang2018spectral,li2019local,qi2020spectral,zheng2021sparse} have also been proposed to introduce related spatial and spectral information in HSI to improve the unmixing results.

In recent years, thanks to the powerful learning and data fitting capabilities of deep neural networks, deep learning-based unmixing methods have attracted more attention. Various deep networks for HU have been proposed, mainly based on deep autoencoder (AE) networks and its variants, which have been proven to be effective for unmixing \cite{licciardi2011pixel}. As a baseline network for unmixing, AE maps the spectrum of mixed pixel into abundance coefficients, and then the decoder reconstructs the input mixed spectrum adopting a linear mapping, which is the endmembers. By imposing constraints such as ANC and ASC on the abundance representations, an AE can be effectively trained \cite{palsson2018hyperspectral}.

EndNet \cite{ozkan2019endnet}, DAEN \cite{su2019daen}, DeepGUn \cite{borsoi2019deep}, uDAS \cite{qu2018udas}, TANet \cite{jin2021tanet}, JMnet \cite{min2021jmnet} and SNMF-Net \cite{xiong2021snmf} are typical examples of such unmixing methods. EndNet proposes a two-staged AE network, and a new loss function composing a Kullback-Leibler divergence term with SAD similarity and a sparsity term. While improving the accuracy of endmember and abundance estimation, the complexity of loss function makes parameter selection a challenging problem. DAEN consists of two parts: the first part employs stacked AEs to initialize the unmixing process, and the latter part uses a variational AE (VAE) to unmix HSI. While in DeepGUn, VAE is used for the endmember generation process. uDAS presents an untied AE network, which uses a denoising constraint and a sparsity constraint on the decoder and encoder, respectively. With the same untied-weighted AE, TANet proposes a two-stream network and introduces an additional network to deal with the problem in the endmembers. Similarly, JMnet introduces an additional discriminator to calculate the Wasserstein distance and introduces a feature matching regularization term. From a different perspective, SNMF-Net proposes an end-to-end unmixing network based on the LMM, which can take advantages of both model-based and learning-based approaches.

In the above AE-based deep unmixing methods, the spatial-contextual information in HSI is ignored. Reviewing traditional unmixing methods based on geometry, statistics and sparse representation, it can be found that most of the unmixing methods, including \cite{wang2017spatial,huang2019spectral,he2017total,zhang2018spectral} and \cite{qi2020spectral}, are mainly proposed for exploring the spatial information in HSI. It is very important to introduce spatial information in the process of unmixing \cite{iordache2012total,mei2015equivalent,he2017total,mei2020improving}, and the superiorities of combining with spatial-contextual information for HSI processing have been confirmed in a large number of literatures \cite{wang2021hyperspectral,zhang2021ssr}.

In the AE-based unmixing networks, the training process based on a single mixed pixel ignores the spatial-contextual information. Therefore, the patchwise or cubewise convolutional neural network (CNN) is introduced to exploit the spatial-contextual information in HSI. Based on the AE architecture presented in \cite{palsson2018hyperspectral}, a structure of multiple parallel AEs using multitask learning is proposed in \cite{palsson2019spectral}, which introduces spatial information into unmixing in the form of HSI patches. In \cite{palsson2021convolutional}, a CNN-based AE model is proposed, which uses HSI patches to train the unmixing network to directly utilize the spatial information of HSI. Most recently, a cycle-consistency AE network based on the perception mechanism is presented in \cite{gao2021cycu}, introducing a new self-perception loss. In \cite{hong2021endmember}, a two-stream siamese unmixing network is proposed, which introduces an additional network to guide the endmember information. Both of these networks can be used to model the spatial-contextual information for spatial-spectral unmixing by adjusting the convolution filter sizes. In \cite{rasti2021undip}, the deep image prior is introduced into unmixing network, which uses CNN to estimate abundance coefficients and incorporates global spatial information. Differently, a spatial-spectral AE network is proposed in \cite{huang2020spatial}, using spatial and spectral information respectively. However, the AE networks are independent, which limits the further improvement of unmixing.

Based on the above analysis, it can be found that the patchwise methods can improve the estimation accuracy of endmembers, because it conforms to the idea of endmember bundles. However, it may reduce the accuracy of abundance estimation and produce blur effects. It is explained in \cite{palsson2021convolutional} and \cite{rasti2021undip} that this is due to the fact that spatial structure information contained in small patches is insufficient to make the convolutional operation perform better. Introducing the spatial information into unmixing process and combining spectral information for effective spatial-spectral unmixing is an important topic in unmixing task. How to efficiently mine the spatial-spectral information in HSI to serve high quality unmixing is the key to improve unmixing accuracy. Therefore, in this article, we explore the efficient utilization of spatial-spectral information in the AE-based deep network. Meanwhile, we propose a novel spatial-spectral collaborative unmixing network, called SSCU-Net, including a new spatial AE network and a spectral AE network. It conducts collaborative training on endmembers and abundances in an end-to-end manner, and greatly improves the performance of unmixing. The main contributions of this article are the following threefold, which distinguish it from other deep unmixing networks.

\emph{1)} We make an important discovery that spatial and spectral information have different influence on endmember extraction and abundance estimation processes in the AE-based unmixing networks. Specifically, the utilization of spatial information has a better effect on the extraction of endmembers, while the usage of spectral information can improve the estimation of abundances. This has an important guiding significance for the application of spatial and spectral information in unmixing networks.

\emph{2)} Based on the discovery, we propose a new spatial-spectral collaborative unmixing network to make appropriate use of spatial-spectral information. It is a two-stream deep network that includes a spatial AE network and a spectral convolutional AE network. Aiming at the particularity of AE-based unmixing network, we design a collaborative strategy for the proposed two-stream AE network to perform high-quality estimation of endmembers and abundances. The collaborative strategy consists of a collaborative loss for abundance fractions and a weight-sharing alternating training strategy for endmember information, which efficiently trains the network in an end-to-end manner.

\emph{3)} In the two-stream deep unmixing network, we propose a new AE network employing spatial information. By developing a superpixel segmentation method based on abundance information, we propose a new superpixel utilization method that integrates superpixels into the AE network. It introduces spatial information into the unmixing process in a more general form to enhance the utilization of spatial information. Whereas the spectral convolutional AE network, with previous developments, adopts a form of spectral convolution to make full use of spectral information.

The rest of this article is structured as follows. Section II introduces the background of AE-based unmixing theory. In Section III, we introduce the proposed SSCU-Net structure. Experimental results and detailed comparisons with state-of-the-art approaches are shown in Section IV. Finally, conclusions are drawn in Section V.

\begin{figure*}[t]
\begin{center}
\includegraphics[width=1\linewidth]{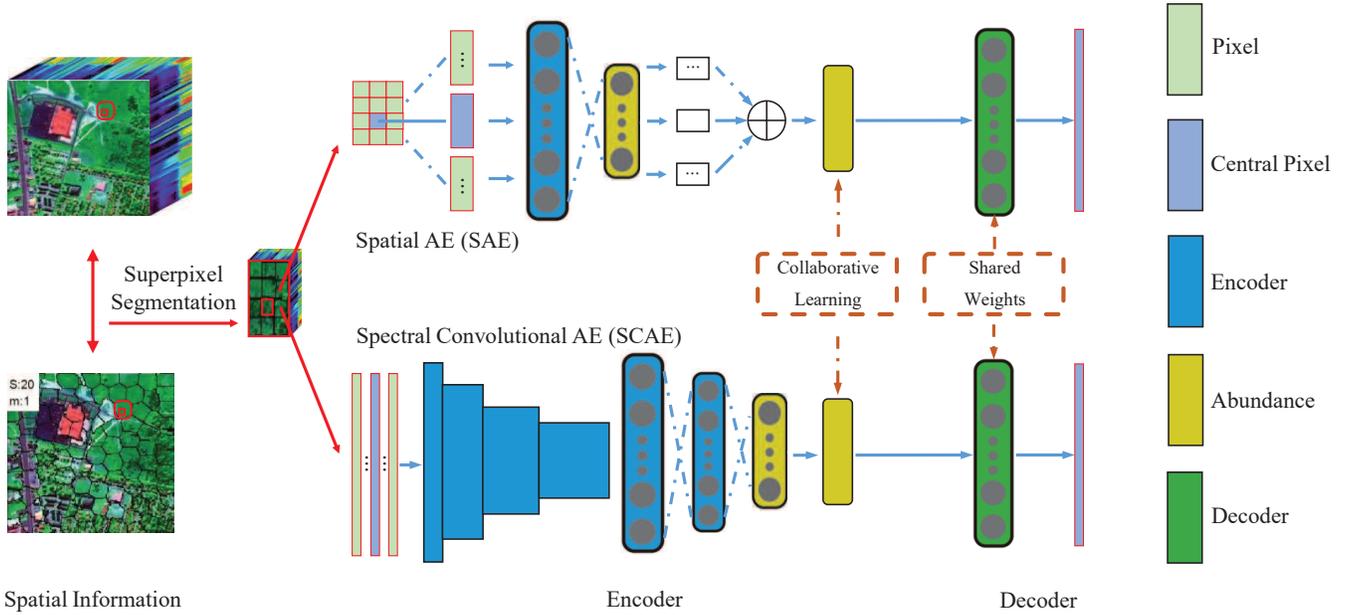}
\end{center}
   \caption{Network structure of the proposed SSCU-Net.}
\label{fig:toy}
\end{figure*}

\section{AE-Based Unmixing Model}
The LMM can be expressed as
\begin{equation}
{Y = AX + N}
\end{equation}
where $Y \in {R^{l \times n}}$ is an observed mixed pixel matrix with $l$ bands and $n$ pixels. $A \in {R^{l \times p}}$ denotes the endmember matrix with $p$ endmember signatures, and $X \in {R^{p \times n}}$ is the corresponding abundance matrix. $N \in {R^{l \times n}}$ represents an additive noise matrix. Subject to physical constraints, the abundance vectors ${x_j}$ need to satisfy the following ANC and ASC constraints:
\begin{equation}
{\left\{ {\begin{array}{*{20}{l}}
{{x_j} \ge 0}\\
{\sum\limits_{i = 1}^p {{x_{ij}} = 1} }
\end{array}} \right.}
\end{equation}

The problem studied in this article is to employ AEs to simultaneously estimate the endmember matrix $A$ and abundance matrix $X$ in (1) via an unsupervised manner. An AE generally consists of an encoder and a decoder.

\emph{1) Encoder:} An encoder uses some trainable network parameters to encode the input spectra $\left\{ {{y_i}} \right\}_{i = 1}^n \in {R^l}$ to a hidden representation ${h_i}$, which is given as
\begin{equation}
{{h_i} = {f_E}({y_i}) = f({W^{(e)T}}{y_i} + {b^{(e)}})}
\end{equation}
where $f$ represents the activation function, and ${W^{(e)}}$ denotes the weight in the $e$th encoder layer, with ${b^{(e)}}$ being the bias. The most widely used activation functions include the rectified linear unit (ReLU), sigmoid, or leaky ReLU functions.

\emph{2) Decoder:} A decoder reconstructs the input spectra using ${h_i}$ through a hidden layer. Based on the LMM, it can be expressed as
\begin{equation}
{{\widehat y_i} = {f_D}({h_i}) = {W^{(d)T}}{h_i}}
\end{equation}
where ${\widehat y_i}$ represents the reconstructed spectrum, ${W^{(d)}}$ is the weight matrix of decoder, which includes the weights between the hidden and output layer.

The AE network learns the representations and weights to reconstruct the spectrum by minimizing the mean square error (MSE)-based objective function, which is defined as
\begin{equation}
{{L_{AE}} = \frac{1}{n}\sum\limits_{i = 1}^n {{{\left\| {{{\widehat y}_i} - {y_i}} \right\|}^2}} }
\end{equation}

It should be noted that in the AE-based unmixing model, we generally do not use the bias, since it often leads to a large negative value \cite{qu2018udas}. Meanwhile, based on equation (4), we can find that the AE model naturally matches with the unmixing problem, that is, the output of encoder can be regarded as the estimated abundances, i.e., ${\widehat x_i} \leftarrow {h_i}$, and the weight of decoder can be considered as the extracted endmembers, i.e., $\widehat A \leftarrow {W^{(d)}}$.

\section{Spatial-Spectral Collaborative Unmixing Network}
In order to efficiently utilize the spatial and spectral information in HSI different from the existing 3D convolutional network structure, we propose a new spatial-spectral collaborative unmixing network. It is a two-stream network that includes a spatial information-oriented spatial AE network and a spectral information-oriented spectral convolutional AE network. The two networks perform endmember extraction and abundance estimation processes in a collaborative manner. Fig.~\ref{fig:toy} illustrates the proposed SSCU-Net architecture.

In the spatial AE network, referred to as SAE, we propose an efficient spatial information processing method and integrate it into the AE network. In the traditional unmixing methods, we observe that the region-based spatial information processing methods have achieved superior results \cite{zortea2009spatial,wang2017spatial,xu2018regional,qi2019region,li2021Superpixel}, among which the superpixel segmentation method is a typical method. Based on this, we propose an abundance-based superpixel segmentation method for processing spatial information. Compared with traditional spatial information processing methods using a sliding window as in \cite{iordache2012total,qi2020spectral,huang2020spatial}, it is a more general form of spatial processing. Meanwhile, the processing form based on abundance is a new way of utilizing spatial information, which can not only make the segmentation results more suitable for the input requirements of AE network, but also reduce the computational complexity. Combining it with the AE unmixing network can make efficient use of spatial information in HSI.

In the spectral convolutional AE network, SCAE for short, based on the previous development, we adopt a 1D spectral convolutional network to make full use of spectral information in HSI. 1D spectral convolution has been widely used in supervised and unsupervised unmixing networks \cite{zhang2018hyperspectral,qi2020deep,khajehrayeni2020hyperspectral}. Convolutional networks have powerful feature extraction capabilities, while HSI has a wealth of continuous spectral information. Therefore, the application of 1D convolutional operations on spectral dimension can effectively extract spectral characteristics.

It is worth pointing out that both SAE and SCAE are unmixing networks based on AE, where SAE mainly explores spatial information and SCAE explores spectral information. They can be used independently as an unmixing network, and the endmember and abundance matrices can be obtained simultaneously in the end. However, we find that the endmember and abundance matrices obtained by the two unmixing networks are different: the accuracy of endmembers extracted by the SAE is better than that of SCAE, while in the case of giving the same endmember matrix, the abundance fractions estimated by the SCAE is better than that of SAE. In other words, the SAE network is more useful to extract endmembers, while the SCAE network is more helpful with abundance estimation. More detailed experimental instructions are described in Section IV-C. This can also be explained in terms of its network structure and the way of using spatial and spectral information. In the SAE network, the spatial information of HSI is mainly explored and a mixed pixel should be input simultaneously with its surrounding pixels. This is consistent with the idea of endmember bundles \cite{jin2021tanet,rasti2021undip}, which is beneficent for the extraction of endmembers. In contrast, its estimation of the abundance fractions of a single pixel is not satisfactory. On the other hand, the SCAE network mainly utilizes the spectral information in HSI and adopts a single sequential input way for the mixed pixels, which will have a better effect on the abundance estimation. This supports the idea of using convolutional networks for abundance estimation in the literature \cite{zhang2018hyperspectral,qi2020deep}.

Based on this discovery, we propose a collaborative strategy to make full use of the respective advantages of SAE and SCAE in the two-stream network. It includes a collaborative loss between abundances and a weight sharing strategy between endmembers, so that they can promote each other, alternate optimization, and finally converge to an optimal solution together. It can not only improve the accuracy of endmember extraction and abundance estimation, but also greatly increase the convergence speed of networks.

In the following subsections, we will describe the components of the proposed SSCU-Net architecture in detail.

\subsection{Spatial AE Network}
Considering Refs. \cite{eches2011enhancing,giampouras2016simultaneously}, the spectral characteristics of adjacent mixed pixels are highly correlated and generally share the same set of endmembers, that is, the same substance is always continuously and smoothly distributed in HSI. Ref. \cite{huang2020spatial} considers spatial information by means of a square sliding window. However, the size and shape of the image patch obtained in this way are fixed, and the spatial and spectral information in local regions cannot be fully utilized. In \cite{huang2020spatial}, we can observe that when the size of the sliding window is larger, the unmixing performance begins to decrease. This is due to the fact that a larger sliding window is more likely to contain pixels that are not similar in abundance to the center pixel.

Hence, in order to obtain more accurate homogeneous regions, we introduce the superpixel segmentation method to obtain local homogeneous regions with varied size and shape, and further explore the role and effect of spatial information in AE-based unmixing. Superpixel segmentation is the mainstream method in image segmentation. It is also widely used in HSI unmixing. Refs. \cite{xu2018regional,li2021Superpixel,li2018superpixel} introduce the spatial information into unmixing more effectively through superpixel segmentation methods. Compared with traditional sliding window-based methods, superpixel segmentation can adaptively partition an image into different sizes and shapes according to data homogeneity, and it shows more competitive performance in terms of noise and outlier suppression, and computational efficiency \cite{jin2021tanet}. In this paper, we adopt the well-known simple linear iterative clustering (SLIC) \cite{achanta2012slic} segmentation method to obtain suitable segmentation results for AE-based unmixing.

In simple terms, for the RGB image in the CIELAB space, the SLIC algorithm clusters pixels according to their color channels and coordinates to effectively generate compact and uniform superpixels. In HSI unmixing, as in \cite{xu2018regional,li2021Superpixel,li2018superpixel}, replacing the luminosity and color component used in RGB images by the spectral vector in HSI, SLIC can be extended to HSI to perform superpixel segmentation. Differently, in our approach, we replace the spectral vector with abundance information. Let $D(i,j)$ denote the dissimilarity of two pixels $i$ and $j$, and it is obtained by calculating the spatial distance ${d_{spa}}(i,j)$ and the abundance distance ${d_{abu}}(i,j)$. Finally, the definition of the distance metrics used in our approach is
\begin{equation}
{D(i,j) = \sqrt {d_{abu}^2(i,j) + {{(\frac{{{d_{spa}}(i,j)}}{S})}^2}{m^2}} }
\end{equation}
\begin{equation}
{{d_{spa}}(i,j) = \sqrt {{{(a - b)}^2} + {{(c - d)}^2}} }
\end{equation}
\begin{equation}
{{d_{abu}}(i,j) = \left\| {{x_i} - {x_j}} \right\|_2^2}
\end{equation}
where $(a,b)$ and $(c,d)$ represent the coordinates of pixel $i$ and $j$, respectively. ${{x_i}}$, ${{x_j}}$ are the abundance vectors. $S = \sqrt {n/Q} $ denotes the nominal size of a superpixel, where $Q$ controls the number of superpixels. And $m$ is a balancing weight controlling the relative importance between abundance and spatial similarity. $S$ and $m$ are two very important parameters. When the nominal size $S$ increases, the obtained superpixel block becomes larger, and as the compact factor $m$ increases, the obtained superpixel block becomes more regular. It is foreseeable that as $m$ tends to infinity, the segmentation results will approach the situation of a square sliding window. In this sense, the window-based methods can be regarded as a special case of superpixel segmentation. This article explores a more general form of spatial information processing by analyzing the parameters $S$ and $m$. Section IV-B gives a more detailed analysis of the influences of $S$ and $m$ on HSI segmentation.

In addition, compared with the previous superpixel segmentation methods based on spectral vector, the introduction of abundance information into superpixel segmentation has two advantages. Firstly, the segmentation results can be more suitable for the input requirements of AE network. In the encoder, we expect to obtain a more accurate hidden representation, i.e., a more accurate abundance. Obviously, it is more direct and accurate to obtain superpixel blocks through abundance information (although relatively rough) than using the original spectral information. Secondly, it can reduce computational complexity. Each mixed pixel in HSI usually contains hundreds of bands, while the dimension of abundance vector of each pixel is generally a single digit, which can greatly reduce the amount of calculation.

After performing superpixel segmentation on the whole HSI, $Q$ superpixel blocks can be obtained. For any superpixel block $SP$, we define its center pixel as ${y_c}$, which is obtained by comparing with the mean value of the pixel coordinates in $SP$. Its definition is given as
\begin{equation}
{{y_c} = \mathop {\arg \min }\limits_{{y_k} \in SP} {d_{spa}}({y_k},{y_v})}
\end{equation}
where ${y_v}$ is a virtual pixel, and its spatial coordinate is the mean value of the pixel coordinates in $SP$.

In the SAE network, rather than using the average value of pixels in a superpixel block as input in unsupervised unmixing methods, we input all pixels in a superpixel block into SAE network in a parallel way, as shown in the top half of Fig.~\ref{fig:toy}. Then the average value of its hidden representations is taken as the hidden representation of the center pixel. The SAE is designed to be a three-layer network, where the encoder is used for dimensional reduction of parallel input pixels, and the decoder is used to reconstruct the spectrum of the center pixel. Drawing on Refs. \cite{qu2018udas} and \cite{min2021jmnet}, the adopted encoder is a single-layer fully connected structure. Studies have shown that deepening the coding layers of the unmixing network may not obtain better endmember extraction results \cite{palsson2018hyperspectral}, which is related to the small amount of data and lack of diversity in HSI, and the usage of more parameters may lead to overfitting. In addition, SSCU-Net aims to develop a general framework for unmixing. Using a simple network structure is also conducive to explore the potential of SSCU-Net.

\begin{table}[t]
  \centering
    \caption{Architecture of the SCAE Network}\label{architecture}
    \scalebox{0.65}{
  \begin{tabular}{lcccccccccccc}
  \hline \hline
\multirow{2}*{Layer name}&Input&Conv1&Conv2&Conv3&Conv4&FC1&FC2&FC3&FC4\\
&&Maxpool&Maxpool&Maxpool&Maxpool\\
\hline
\multirow{2}*{Kernel size}&$1 \times l$&$1 \times 5$&$1 \times 4$&$1 \times 5$&$1 \times 4$\\
&&$1 \times 2$&$1 \times 2$&$1 \times 2$&$1 \times 2$\\
\hline
\multirow{1}*{Feature map/Units}&1&3&6&12&24&concat&100&$p$&$l$\\
\hline \hline
\end{tabular}}
\end{table}

The ReLU activation function is used in the encoder, and it also contains a BN layer and a Dropout layer. After the encoding operation, the pixels in a superpixel block are compressed into low-dimensional representations, and the transformation can be expressed as
\begin{equation}
{\begin{array}{l}
{h_k} = ReLU(Dropout(BN({W^{(e)}}{y_k}))),\\
\quad \quad \ {y_k} \in SP,k = 1,...,K
\end{array}}
\end{equation}
where ${W^{(e)}}$ is the fully connected weight without bias, and $K$ is the number of mixed pixels in $SP$.

According to equation (3), ${h_k}$ is the estimated abundance vector. To satisfy the ASC constraint, we adopt the ${l_1}$ norm, and the final abundance of each pixel ${y_k}$ in $SP$ is expressed as
\begin{equation}
{{x_k} = \frac{{{h_k}}}{{{{\left\| {{h_k}} \right\|}_1} + \varepsilon }}}
\end{equation}
where $\varepsilon $ is a very small constant to avoid zero-division.

Then, we can obtain the hidden representation of the center pixel ${y_c}$, i.e., its abundance. Considering that superpixels tend to be homogeneous, we take the average abundance of pixels in a superpixel block as the abundance of the center pixel ${y_c}$, i.e.,
\begin{equation}
{{x_c} = \frac{1}{K}\sum\limits_{k = 1}^K {{x_k}} }
\end{equation}

Finally, through the linear decoder, the center pixel ${y_c}$ is reconstructed as
\begin{equation}
{{\widehat y_c} = {W^{(d)}}{x_c}}
\end{equation}
where ${W^{(d)}}$ is the weight matrix of decoder, i.e., the extracted endmember matrix.

Considering the scale invariance of spectral angle distance (SAD), many AE-based researches apply it as the loss function in unmixing \cite{palsson2018hyperspectral,ozkan2019endnet,palsson2021convolutional}. For this reason, we also adopt the SAD as the underlying loss function. Meanwhile, the ${l_{1/2}}$ norm of abundance vector is applied, which can obtain a sparser result than the ${l_1}$ norm \cite{qian2011hyperspectral} to promote the sparsity of the estimated abundance. The objective function of the SAE network is written as
\begin{equation}
{{L_{SAE}} = \frac{1}{Q}\sum\limits_i^Q {\left( {SAD(y_c^i,\widehat y_c^i) + \lambda {{\left\| {x_c^i} \right\|}_{\frac{1}{2}}}} \right)} }
\end{equation}
where ${y_c^i}$ and ${\widehat y_c^i}$ denote the original and reconstruction center pixel in the $i$th superpixel block, respectively, and ${x_c^i}$ represents the abundance vector of ${y_c}$ in the $i$th superpixel block.

\begin{figure}[!t]
\centering
\subfigure[]{
\begin{minipage}{0.40\linewidth}
\centering
\includegraphics[width=\textwidth]{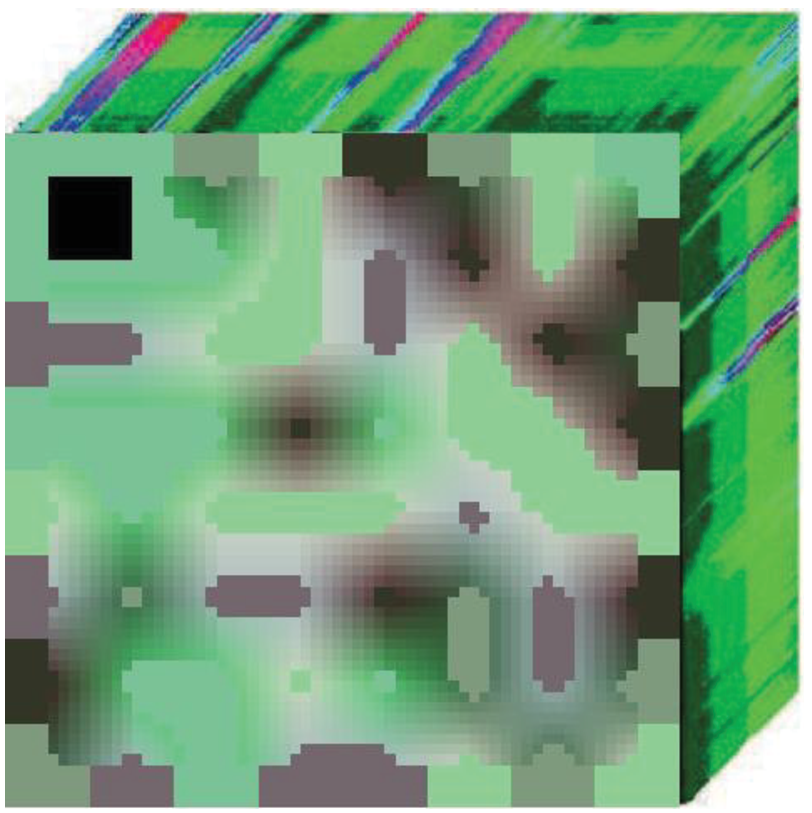}
\end{minipage}
}
\subfigure[]{
\begin{minipage}{0.53\linewidth}
\centering
\includegraphics[width=\textwidth]{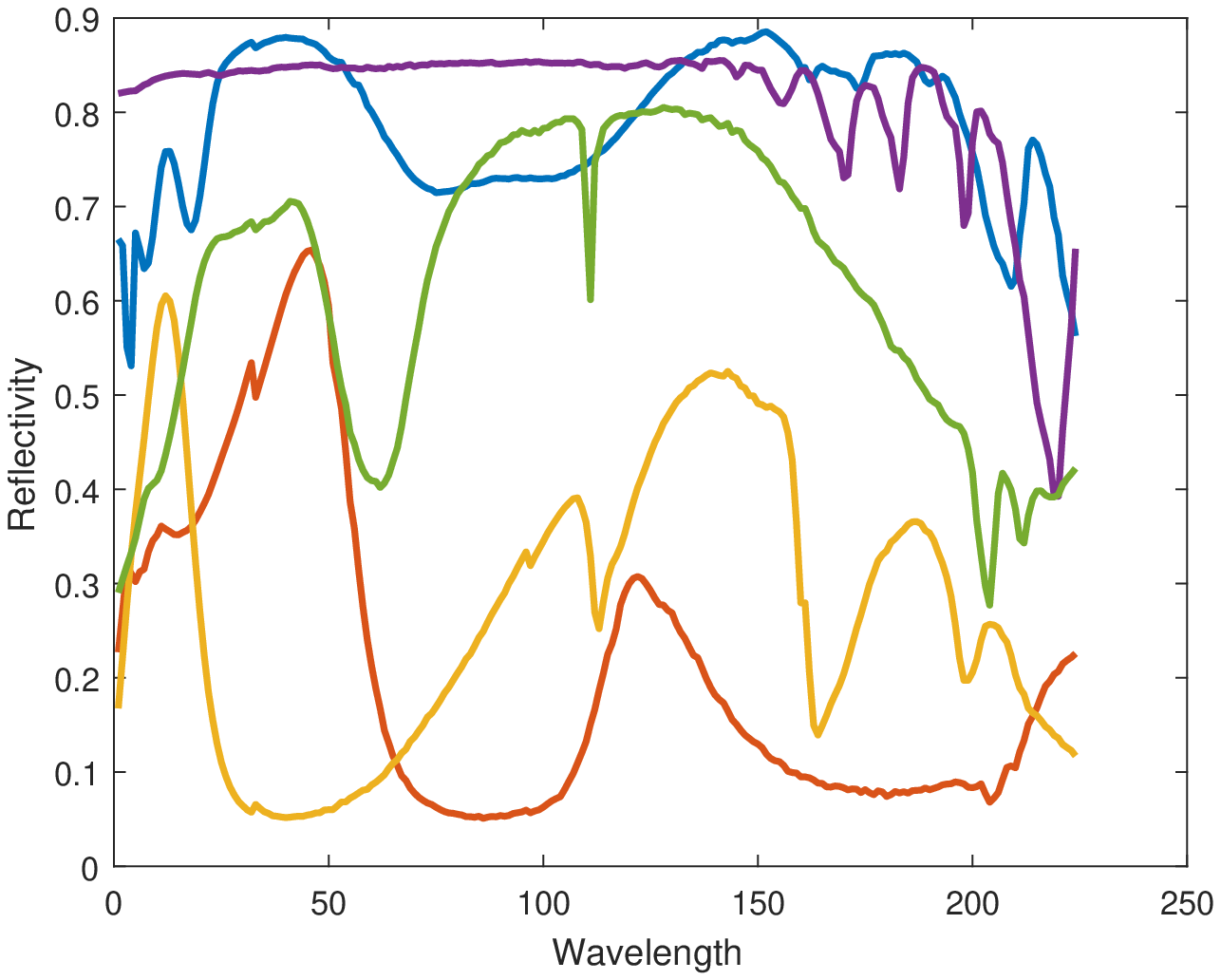}
\end{minipage}
}
\caption{(a) The synthetic data cube. (b) Endmembers in the synthetic data selected from the USGS spectral library.}\label{synthetic}
\end{figure}

\subsection{Spectral Convolutional AE Network}
The abundant spectral information in HSI can benefit further improvement of unmixing performance.
Based on the development in \cite{huang2020spatial}, we employ a 1D spectral convolutional network to take full advantage of the spectral information in HSI.

As shown in the bottom half of Fig.~\ref{fig:toy}, the encoder of the proposed SCAE network adopts a multi-layer spectral convolution, and the decoder is the same as in the SAE network. The network configuration of SCAE is shown in Table \ref{architecture}. Different from the parallel input in the SAE network, the mixed pixels in a superpixel block are serially input to the SCAE network, that is, it performs unmixing pixel by pixel. We define an abundance function $Enc( \cdot )$ and a reconstructed pixel function $Dec( \cdot )$, which respectively correspond to the encoding and decoding operations. Then, the abundance vector ${x_k}$ of pixel ${y_k}$ in any superpixel $SP$ can be estimated by $Enc( \cdot )$ as
\begin{equation}
{{x_k} = Enc({y_k},\theta ){\rm{   }}\quad s.t.\ {\rm{  }}{x_{ki}} \ge 0,{\rm{  }}\sum\limits_i^p {{x_{ki}} = 1} }
\end{equation}
where $\theta $ stands for the training parameters in the encoder, and ${x_{ki}}$ is the $i$th element of ${x_k}$. The ANC and ASC constraints are the same as used in the SAE network. The reconstruction of pixel ${y_k}$ in any superpixel $SP$ can be represented by $Dec( \cdot )$ as
\begin{equation}
{{\widehat y_k} = Dec({x_k},{W^{(d)}}) = {W^{(d)}}{x_k}}
\end{equation}
which is the same as equation (13), and ${W^{(d)}}$ is the endmember matrix.

The objective function of the SCAE network is also consistent with that of the SAE, composed of a reconstruction error term and a sparse term as
\begin{equation}
{{L_{SCAE}} = \frac{1}{Q}\sum\limits_i^Q {\frac{1}{K}\sum\limits_k^K {\left( {SAD(y_k^i,\widehat y_k^i) + \lambda {{\left\| {x_k^i} \right\|}_{\frac{1}{2}}}} \right)} } }
\end{equation}

\begin{figure}[!t]
\centering
\subfigure[]{
\begin{minipage}{0.40\linewidth}
\centering
\includegraphics[width=\textwidth]{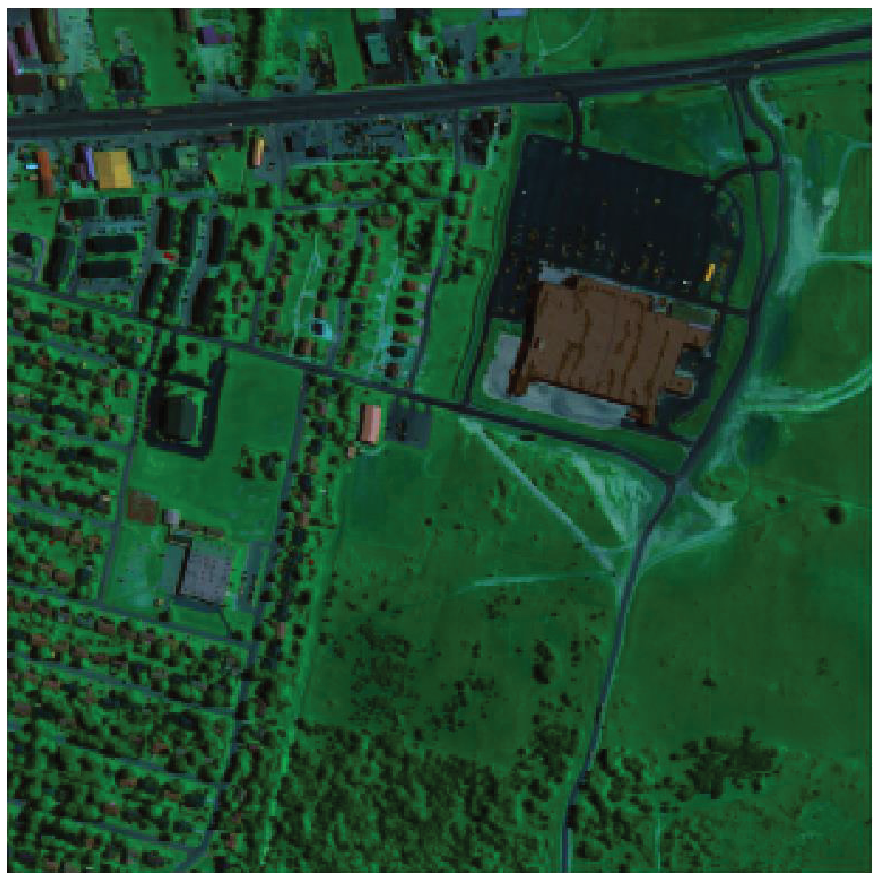}
\end{minipage}
}
\subfigure[]{
\begin{minipage}{0.53\linewidth}
\centering
\includegraphics[width=\textwidth]{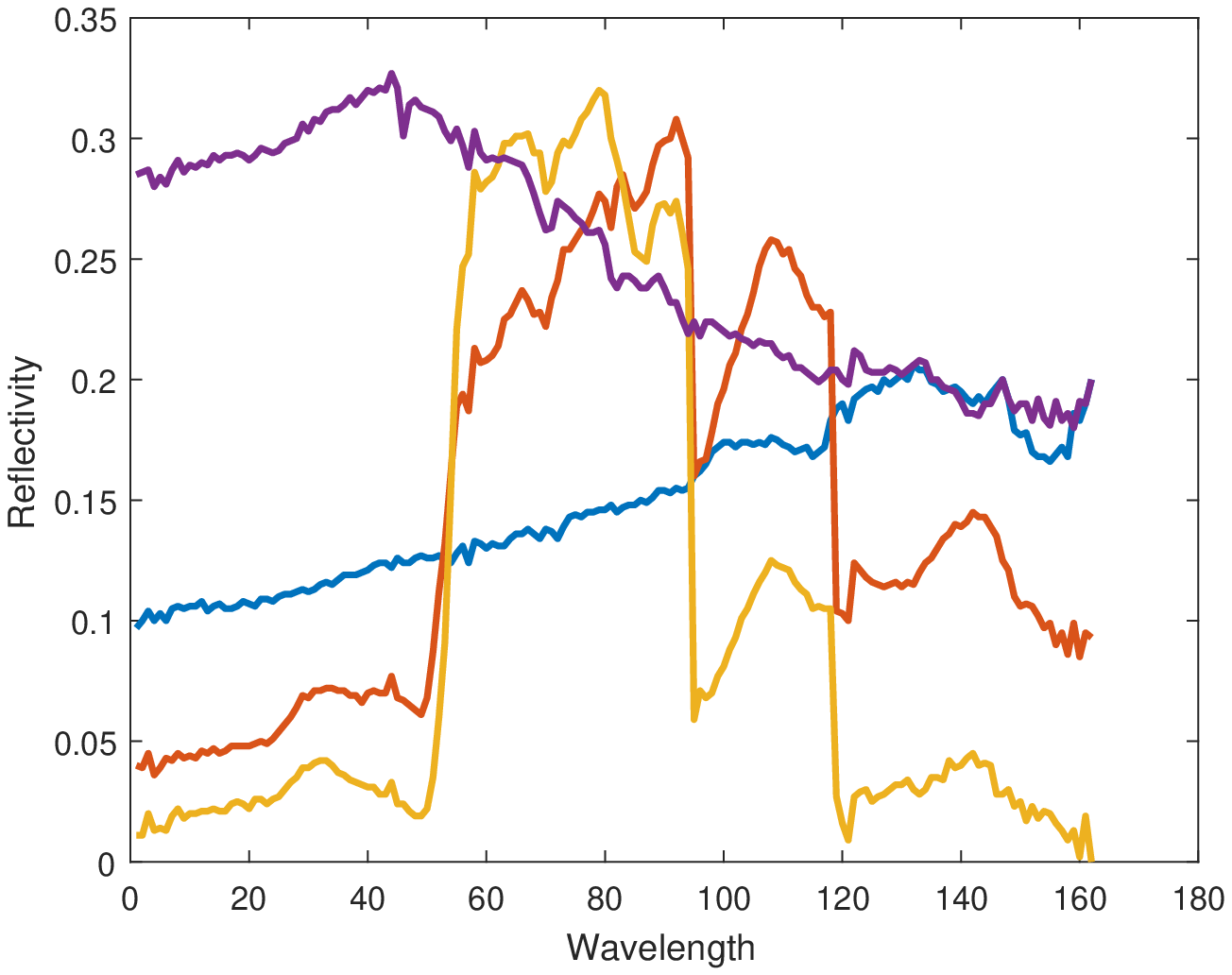}
\end{minipage}
}
\caption{(a) The Urban data set with bands 10, 90, 160. (b) Endmembers in the Urban data set.}\label{urban}
\end{figure}

\subsection{Collaborative Strategy and Objective Functions}
Based on the important discovery, we propose a collaborative strategy to fully explore the potential of the proposed two-stream network. Firstly, the decoders of SAE and SCAE networks share a set of weights. In AE-based unmixing, the weight matrix of decoder represents the estimated endmember matrix. Since the endmembers of SAE and SCAE are estimated from a same HSI, their weights have the same meaning. By sharing the weights, the more accurate endmembers extracted by the SAE network will have a positive impact on the SCAE network, and then SCAE can obtain a more accurate abundance vector.

On the other hand, we design a collaborative term between abundances, which can generate a more accurate abundance by minimizing the abundance vectors estimated by the two networks. It can be expressed as
\begin{equation}
{{L_{{\rm{COL}}}} = \frac{1}{Q}\sum\limits_i^Q {\frac{1}{K}\sum\limits_k^K {\left( {\left\| {x{{_k^i}_{(SAE)}} - x{{_k^i}_{(SCAE)}}} \right\|_2^2} \right)} } }
\end{equation}
By introducing collaborative learning between abundances, the high precision abundance generated by the SCAE network will have a positive impact on the SAE network, and then SAE can obtain more accurate endmembers, forming a closed loop with the weight sharing strategy to obtain high quality endmembers and abundances. In this way, the two-stream network can correct each other toward a more accurate and interpretable result.

It is worth noting that the two-stream network structure is alternately optimized, which can maximize the utilization of collaborative strategy and greatly improve the convergence speed of the network. The final objective function can be expressed as
\begin{equation}
{L = {L_{SAE}} + {L_{SCAE}} + \mu {L_{COL}}}
\end{equation}
where $\mu $ is a tradeoff parameter used to balance the pixel reconstruction term and the abundance collaborative term.

\begin{figure}[!t]
\centering
\subfigure[]{
\begin{minipage}{0.40\linewidth}
\centering
\includegraphics[width=\textwidth]{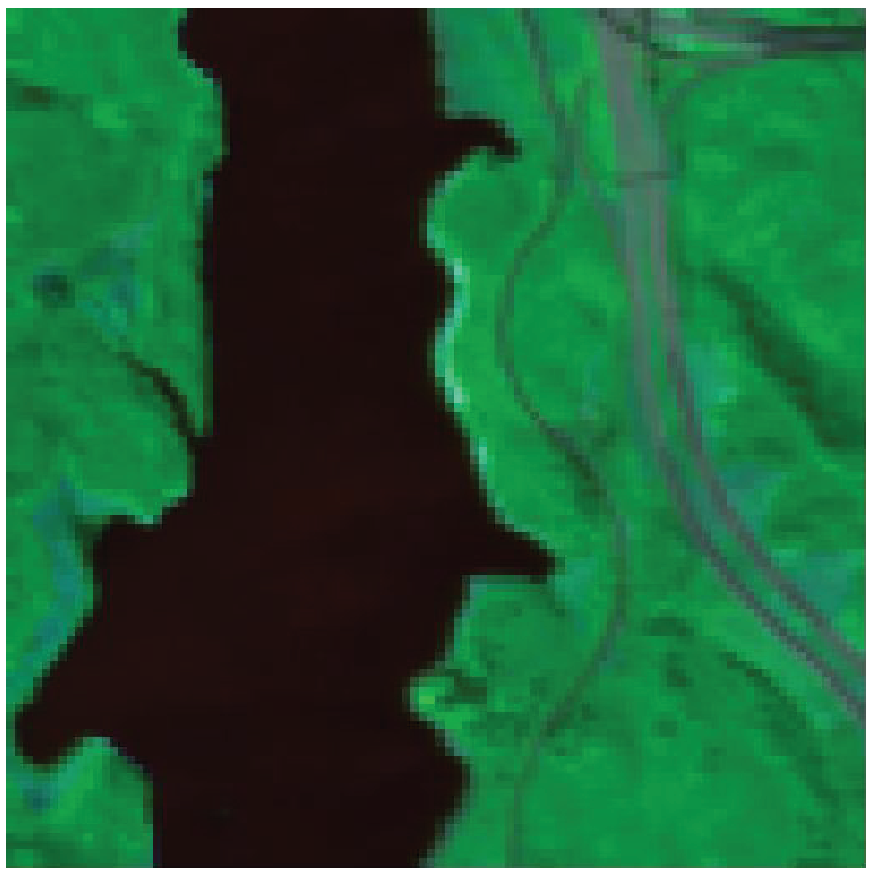}
\end{minipage}
}
\subfigure[]{
\begin{minipage}{0.53\linewidth}
\centering
\includegraphics[width=\textwidth]{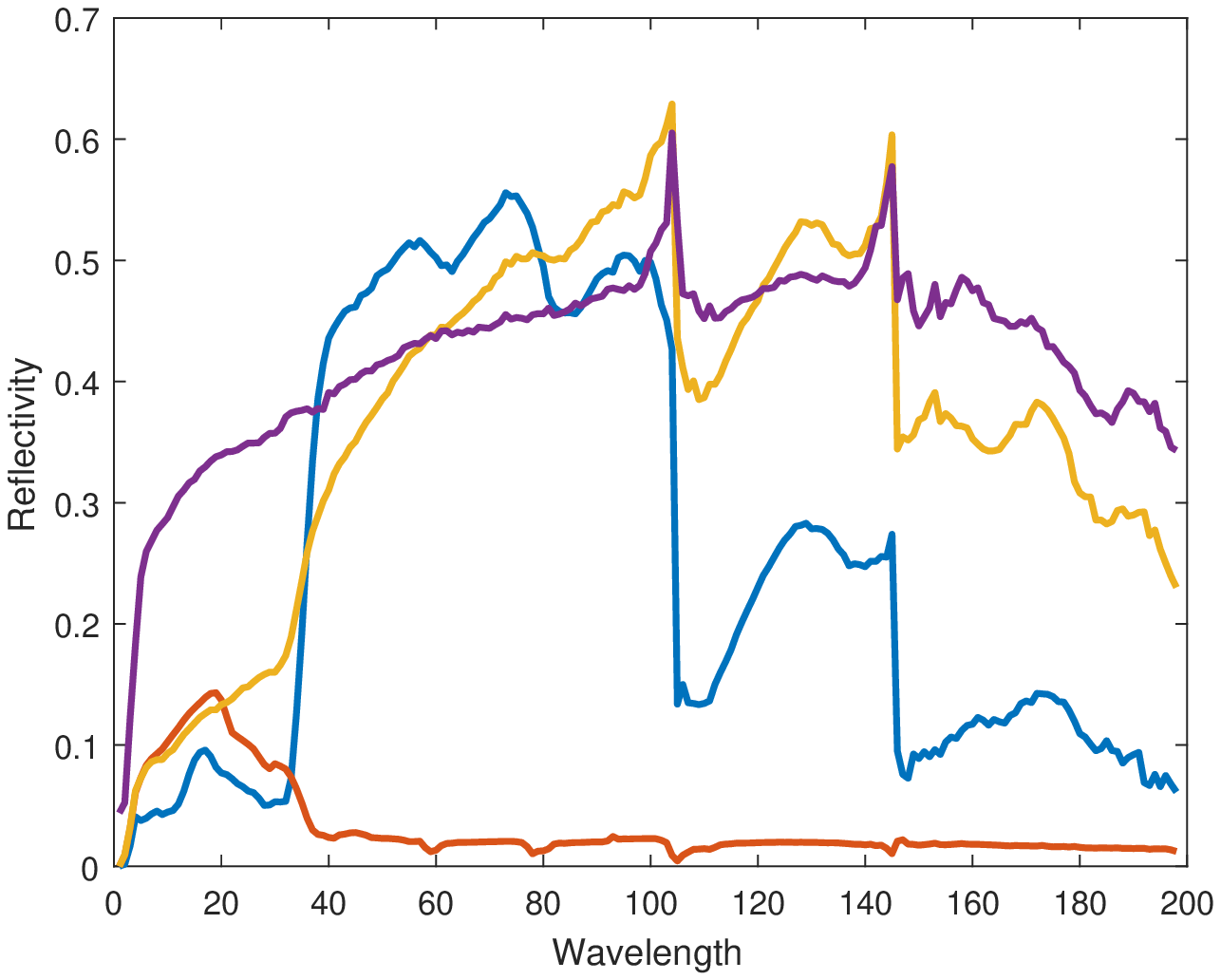}
\end{minipage}
}
\caption{(a) The Jasper Ridge data set with bands 10, 90, 180. (b) Endmembers in the Jasper Ridge data set.}\label{jasper}
\end{figure}

\begin{figure}[!t]
\centering
\subfigure[]{
\begin{minipage}{0.40\linewidth}
\centering
\includegraphics[width=\textwidth]{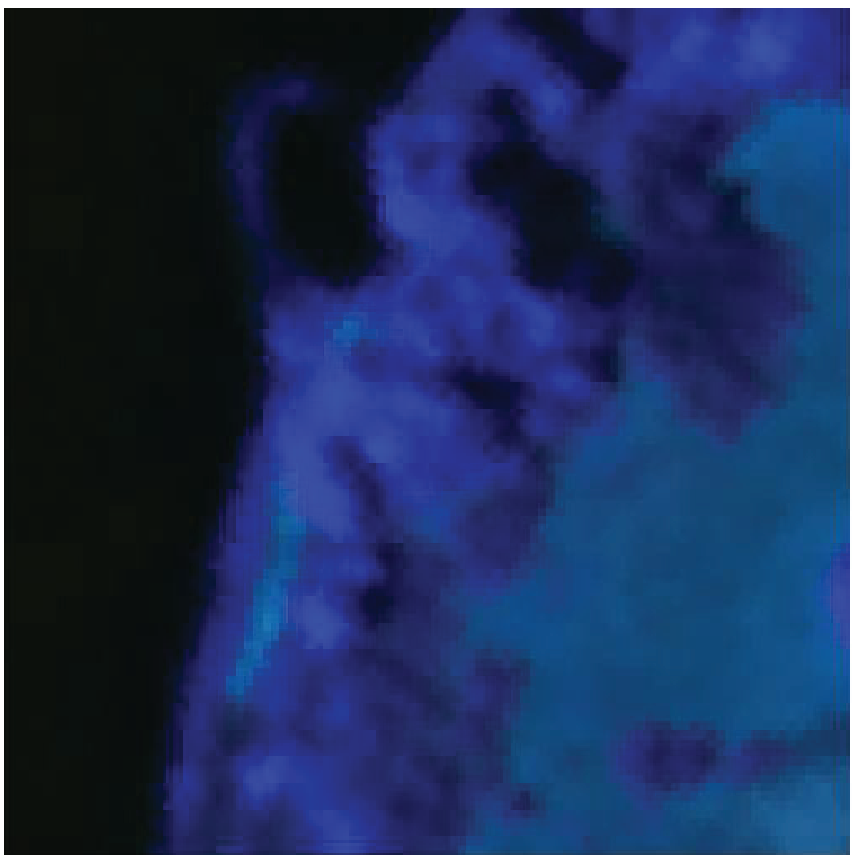}
\end{minipage}
}
\subfigure[]{
\begin{minipage}{0.53\linewidth}
\centering
\includegraphics[width=\textwidth]{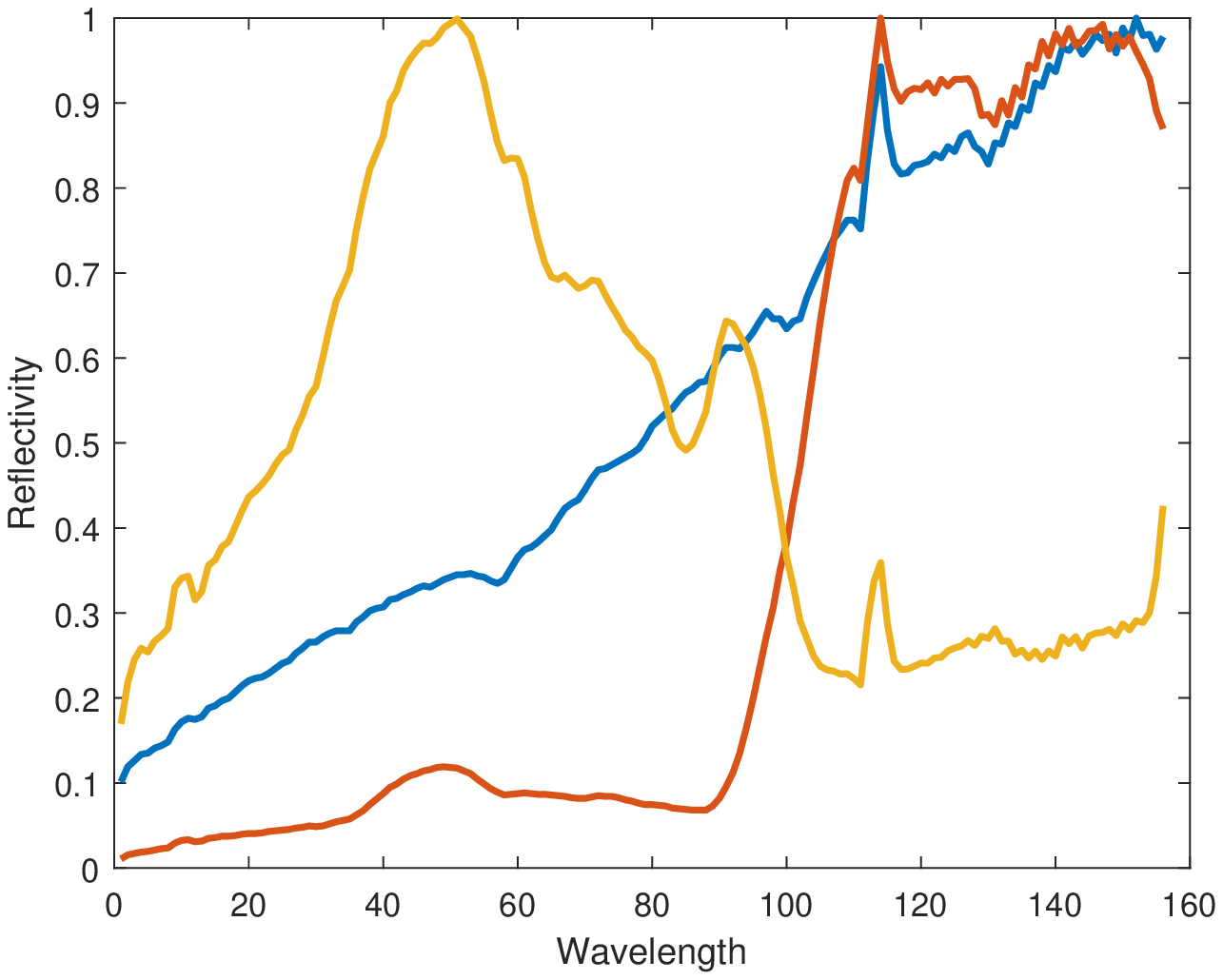}
\end{minipage}
}
\caption{(a) The Samson data set with bands 10, 90, 150. (b) Endmembers in the Samson data set.}\label{samson}
\end{figure}

\section{Experiments}
In this section, we comprehensively explore the effectiveness of the proposed SSCU-Net for unmixing, using synthetic and real hyperspectral data. Seven classical and state-of-the-art unmixing methods are selected for comparison, including VCA \cite{nascimento2005vertex}, ${L_{1/2}}$-NMF \cite{qian2011hyperspectral}, Dgs-NMF \cite{zhu2014spectral}, EndNet \cite{ozkan2019endnet}, TANet \cite{jin2021tanet}, CNNAEU \cite{palsson2021convolutional}, SSAE \cite{huang2020spatial}. Among these comparison methods, VCA is a baseline method, ${L_{1/2}}$-NMF and Dgs-NMF are methods based on NMF, ${L_{1/2}}$-NMF considers the constraint of sparsity, and Dgs-NMF considers spatial-spectral information. EndNet, TANet, CNNAEU, and SSAE are unmixing methods based on AE. EndNet and TANet only use spectral information, while CNNAEU and SSAE are methods based on spatial-spectral information.

\subsection{Hyperspectral Data Sets}
One synthetic data set and two real data sets are used to evaluate the performance of different algorithms. The synthetic data set with different levels of noise being added is used to test the robustness of the algorithms, and the real data sets are used to test the effectiveness of unmixing algorithms in real scenarios.

\begin{figure}[!t]
\centering
\subfigure[]{
\begin{minipage}{0.34\linewidth}
\centering
\includegraphics[width=\textwidth]{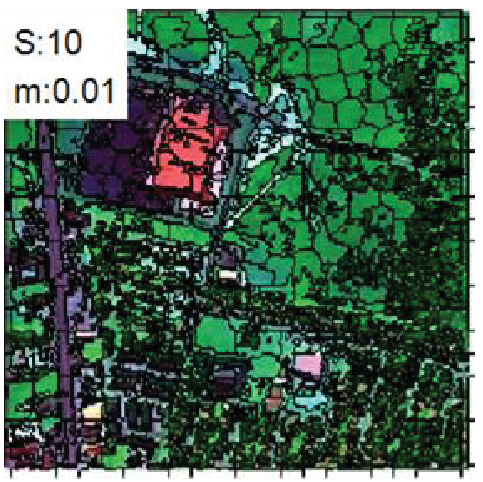}
\end{minipage}
}
\hspace{-0.7cm}
\subfigure[]{
\begin{minipage}{0.34\linewidth}
\centering
\includegraphics[width=\textwidth]{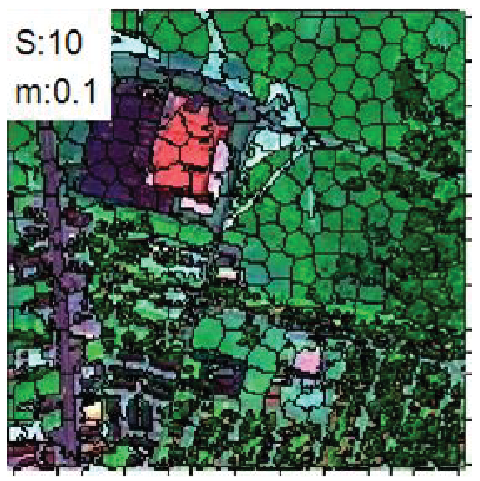}
\end{minipage}
}
\hspace{-0.7cm}
\subfigure[]{
\begin{minipage}{0.34\linewidth}
\centering
\includegraphics[width=\textwidth]{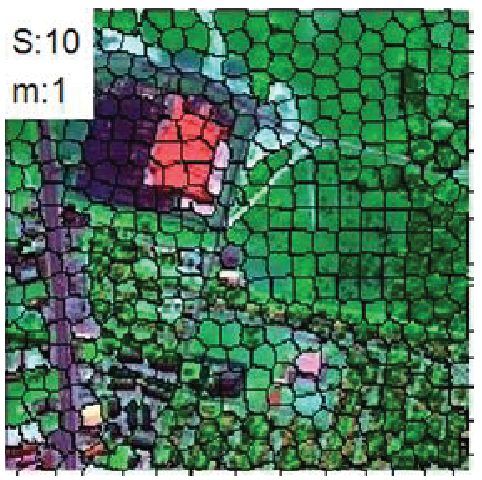}
\end{minipage}
}
\subfigure[]{
\begin{minipage}{0.34\linewidth}
\centering
\includegraphics[width=\textwidth]{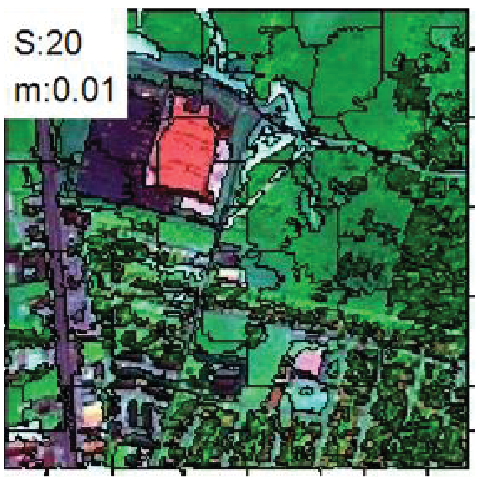}
\end{minipage}
}
\hspace{-0.7cm}
\subfigure[]{
\begin{minipage}{0.34\linewidth}
\centering
\includegraphics[width=\textwidth]{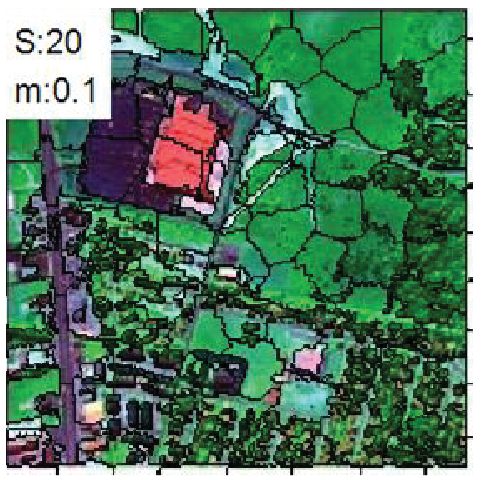}
\end{minipage}
}
\hspace{-0.7cm}
\subfigure[]{
\begin{minipage}{0.34\linewidth}
\centering
\includegraphics[width=\textwidth]{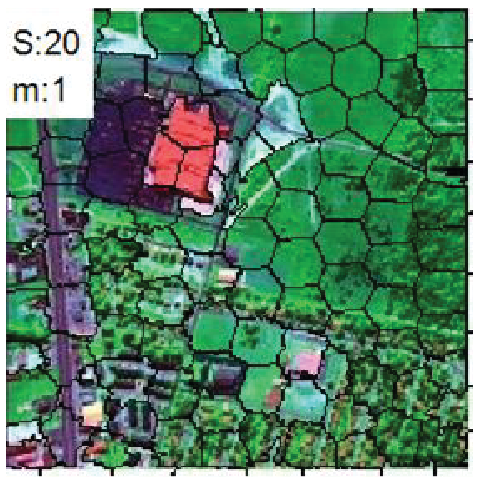}
\end{minipage}
}
\caption{Superpixel segmentation results on the Urban data set under different $S$ and $m$. (a) $S=10$ and $m=0.01$. (b) $S=10$ and $m=0.1$. (c) $S=10$ and $m=1$. (d) $S=20$ and $m=0.01$. (e) $S=20$ and $m=0.1$. (f) $S=20$ and $m=1$.}\label{SP}
\end{figure}

\begin{figure}[!t]
\centering
\subfigure[]{
\begin{minipage}{0.32\linewidth}
\centering
\includegraphics[width=\textwidth]{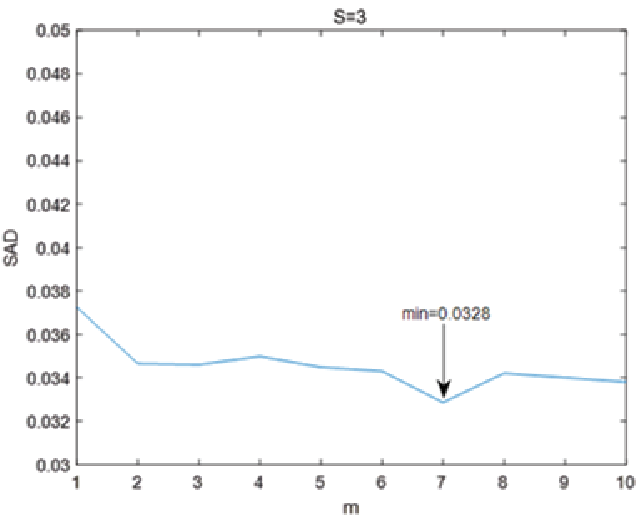}
\end{minipage}
}
\hspace{-0.5cm}
\subfigure[]{
\begin{minipage}{0.32\linewidth}
\centering
\includegraphics[width=\textwidth]{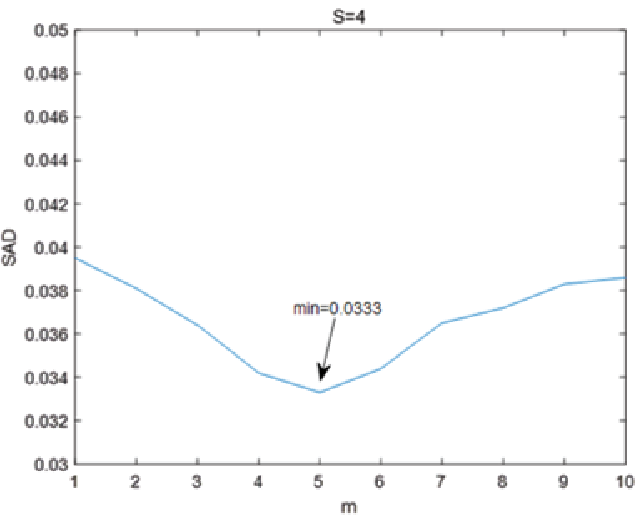}
\end{minipage}
}
\hspace{-0.5cm}
\subfigure[]{
\begin{minipage}{0.32\linewidth}
\centering
\includegraphics[width=\textwidth]{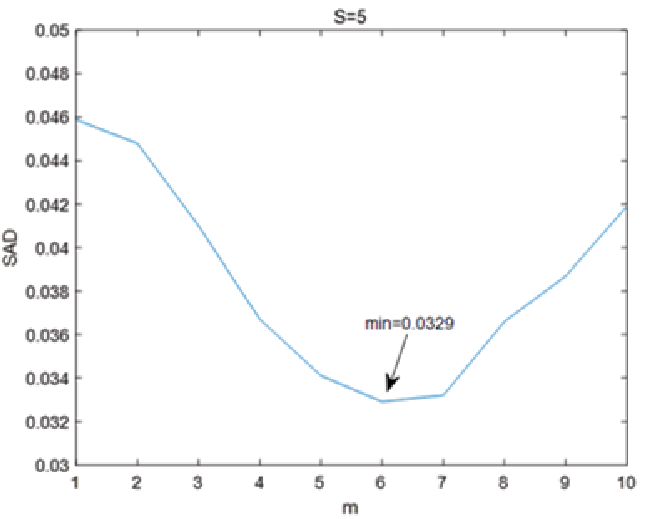}
\end{minipage}
}
\subfigure[]{
\begin{minipage}{0.32\linewidth}
\centering
\includegraphics[width=\textwidth]{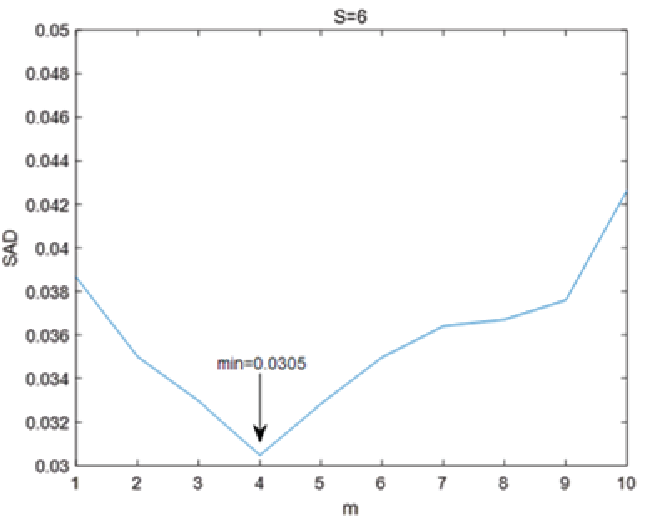}
\end{minipage}
}
\hspace{-0.5cm}
\subfigure[]{
\begin{minipage}{0.32\linewidth}
\centering
\includegraphics[width=\textwidth]{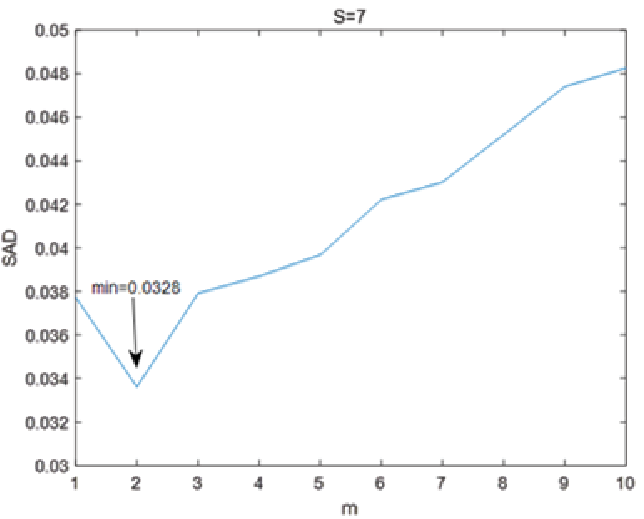}
\end{minipage}
}
\caption{The mean SAD results under different $S$ and $m$ on the Urban data set. (a) $S=3$. (b) $S=4$. (c) $S=5$. (d) $S=6$. (e) $S=7$.}\label{Svsm}
\end{figure}

\emph{1) Synthetic Data Set:} The considered synthetic data set consists of $64 \times 64$ pixels, provided by Tang et al. \cite{tang2015sparse}. The data cube is shown in Fig. \ref{synthetic}, which exhibits a spatial homogeneity. The five endmembers are selected from the U.S. Geological Survey (USGS) spectral library \cite{clark2007usgs}, and the abundances are simulated to satisfy the ANC and ASC constraints. Meanwhile, different levels of independent identically distributed Gaussian noise are added to the data set.

\emph{2) Real Data Set 1 (Urban):} The Urban data set is acquired via the hyperspectral digital image collection experiment (HYDICE) sensor, and it contains 210 spectral bands (0.4-2.5 $\mu $m). As used in \cite{min2021jmnet,palsson2021convolutional}, after removing the contaminated bands, the data set includes $307 \times 307$ pixels and 162 channels. The RGB image of the Urban data set is shown in Fig. \ref{urban}, where four main endmembers are observed: Asphalt, Grass, Tree, Roof.

\emph{3) Real Data Set 2 (Jasper Ridge):} The Jasper Ridge data set is captured by the airborne visible infrared imaging spectrometer (AVIRIS) sensor, and the pixels are recorded on 224 spectral bands (0.38-2.5 $\mu $m). The low signal-to-noise ratio (SNR) and water-vapor absorption bands are eliminated in the subregion of the data used for unmixing. Similar to \cite{gao2021cycu,ozkan2019endnet}, the final data contains $100 \times 100$ pixels and 198 spectral bands. The RGB image of the Jasper Ridge data set is shown in Fig. \ref{jasper}, where four main endmembers are investigated: Tree, Water, Soil and Road.

\emph{4) Real Data Set 3 (Samson):} The Samson data set is collected by the SAMSON sensor, which contains 156 spectral bands (0.4-0.9 $\mu $m). Considering that the original image is large, the final data set includes $95 \times 95$ pixels and 156 channels as in \cite{zhu2014spectral,ozkan2019endnet}. The RGB image of the Samson data set is shown in Fig. \ref{samson}, in which three constituent materials are observed: Soil, Tree, Water.

\begin{figure}[!t]
\centering
\subfigure[]{
\begin{minipage}{0.46\linewidth}
\centering
\includegraphics[width=\textwidth]{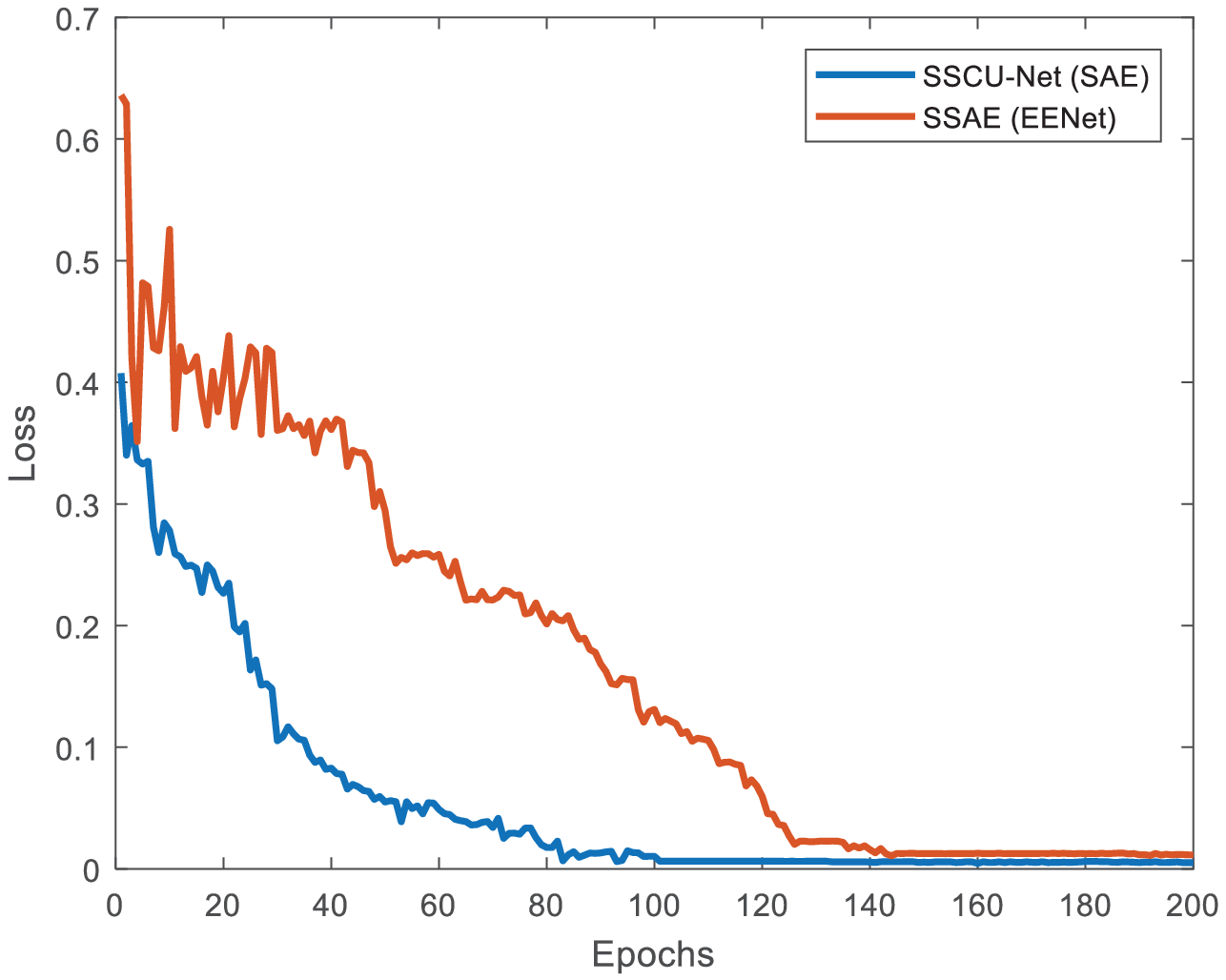}
\end{minipage}
}
\subfigure[]{
\begin{minipage}{0.46\linewidth}
\centering
\includegraphics[width=\textwidth]{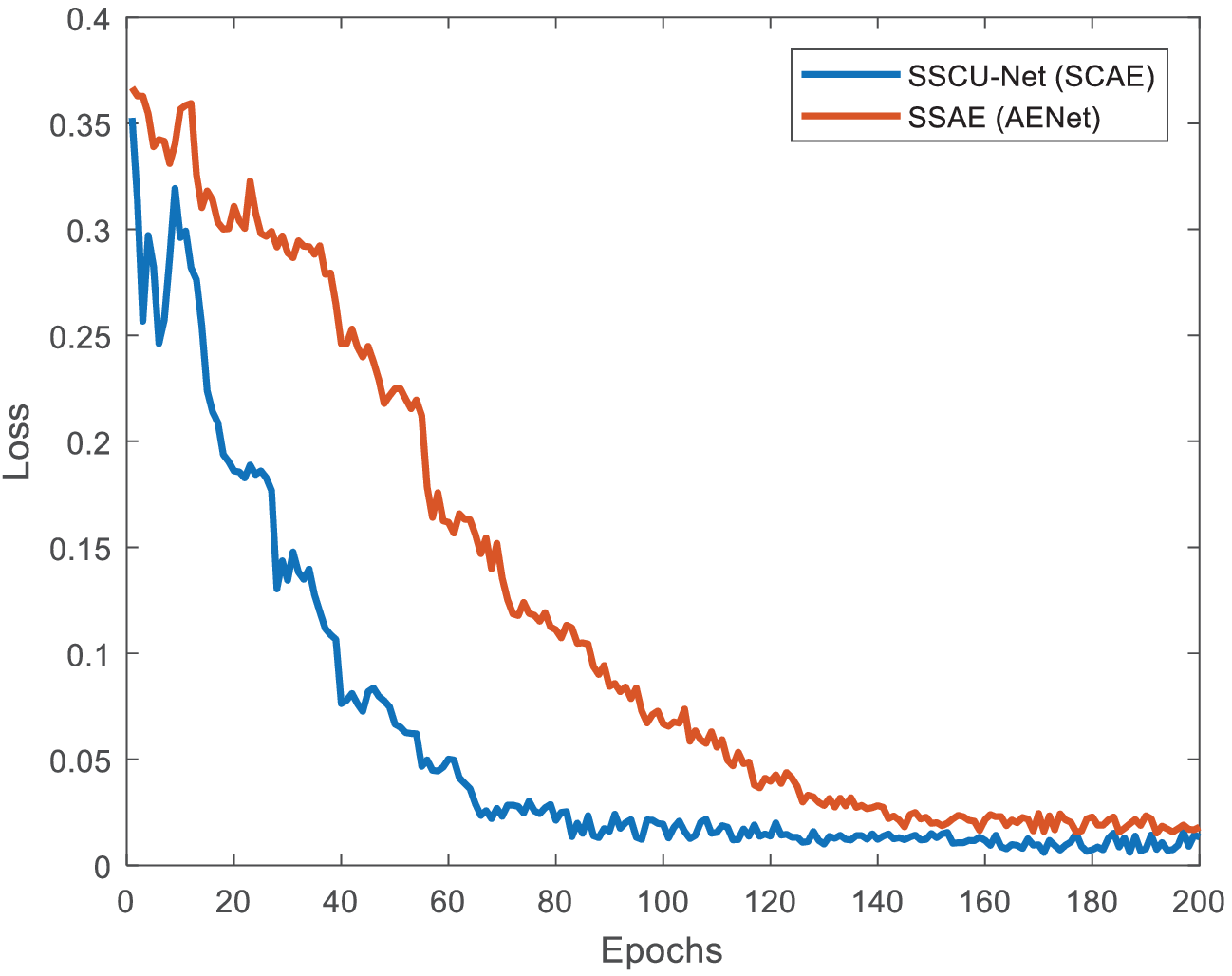}
\end{minipage}
}
\caption{Loss as a function of the number of iterations. (a) The convergence curve for SAE and EENet. (b) The convergence curve for SCAE and AENet.}\label{convergence}
\end{figure}

\subsection{Experimental Settings}
The SAD and the root-mean-square error (RMSE) are used to evaluate the performance of unmixing. SAD calculates the spectral angle between the real endmembers ${a_k}$ and estimated endmembers ${\widehat a_k}$, which is described as
\begin{equation}
{SAD({a_k},{\widehat a_k}) = \arccos \left( {\frac{{{a_k} \cdot {{\widehat a}_k}}}{{{{\left\| {{a_k}} \right\|}_2}{{\left\| {{{\widehat a}_k}} \right\|}_2}}}} \right)}.
\end{equation}

RMSE measures the error between the true abundance vector ${x_k}$ and estimated vector ${\widehat x_k}$, defined as
\begin{equation}
{RMSE({x_k},{\widehat x_k}) = \sqrt {\frac{1}{n}\sum\limits_{k = 1}^n {\left\| {{x_k} - {{\widehat x}_k}} \right\|_2^2} } }.
\end{equation}
Note that the smaller values of the metrics indicate a better performance of the unmixing methods.

Our proposed SSCU-Net model is implemented under the environment of Python 3.6 and PyTorch 1.0 with CPU. We use Adam optimizer to update the parameters. The learning rates of encoder and decoder are set to $1e-4$ and $1e-5$, respectively. Moreover, the model is trained with a maximum number of 200 epochs and a batch size of 128. The trade-off parameters $\lambda $ and $\mu $ are set to $5e-5$ and $5e-1$, respectively. For fair comparisons, the endmember matrix extracted by VCA is used as (partial) initialization for all the algorithms. The abundance information used for superpixel segmentation is obtained through the fully constrained least squares (FCLS) algorithm [6], given the endmembers from VCA.

In addition, computational complexity or memory requirement is an important parameter for unmixing methods based on deep neural networks. Herein, we calculate the amount of parameters for five AE-based unmixing networks. The number of parameters for EndNet and CNNAEU is 1.0K-1.5K, for TANet is about 15K, while for SSAE and SSCU-Net is about 20K. It is worth noting that the structure of current deep unmixing network is relatively simple, and the number of parameters is not large. Under the condition of controlling a certain amount of parameters, SSCU-Net can achieve better unmixing effect.

\subsection{Experiments on SSCU-Net}
\emph{1) Parameter Analysis of Spatial Information:} In the SAE network, thanks to the efficient utilization of spatial information by superpixel blocks, SAE can extract more accurate endmembers. In superpixel segmentation, the parameters $S$ and $m$ control the size, shape and compactness of the superpixel blocks. Taking the Urban data set as an example, Fig. \ref{SP} shows the results of superpixel segmentation. We can see that as the values of $S$ and $m$ changing, the segmentation results are consistent with the analysis in Section III. By increasing the value of $m$, the result of superpixel segmentation is close to the way of a square sliding window. The regular square window cannot perfectly represent the spatial information. On the contrary, the superpixels in Fig. \ref{SP} can naturally express the spatial neighborhoods with adaptively adjusted shape and size of superpixel blocks.

\begin{figure}[!t]
\centering
\subfigure[]{
\begin{minipage}{0.46\linewidth}
\centering
\includegraphics[width=\textwidth]{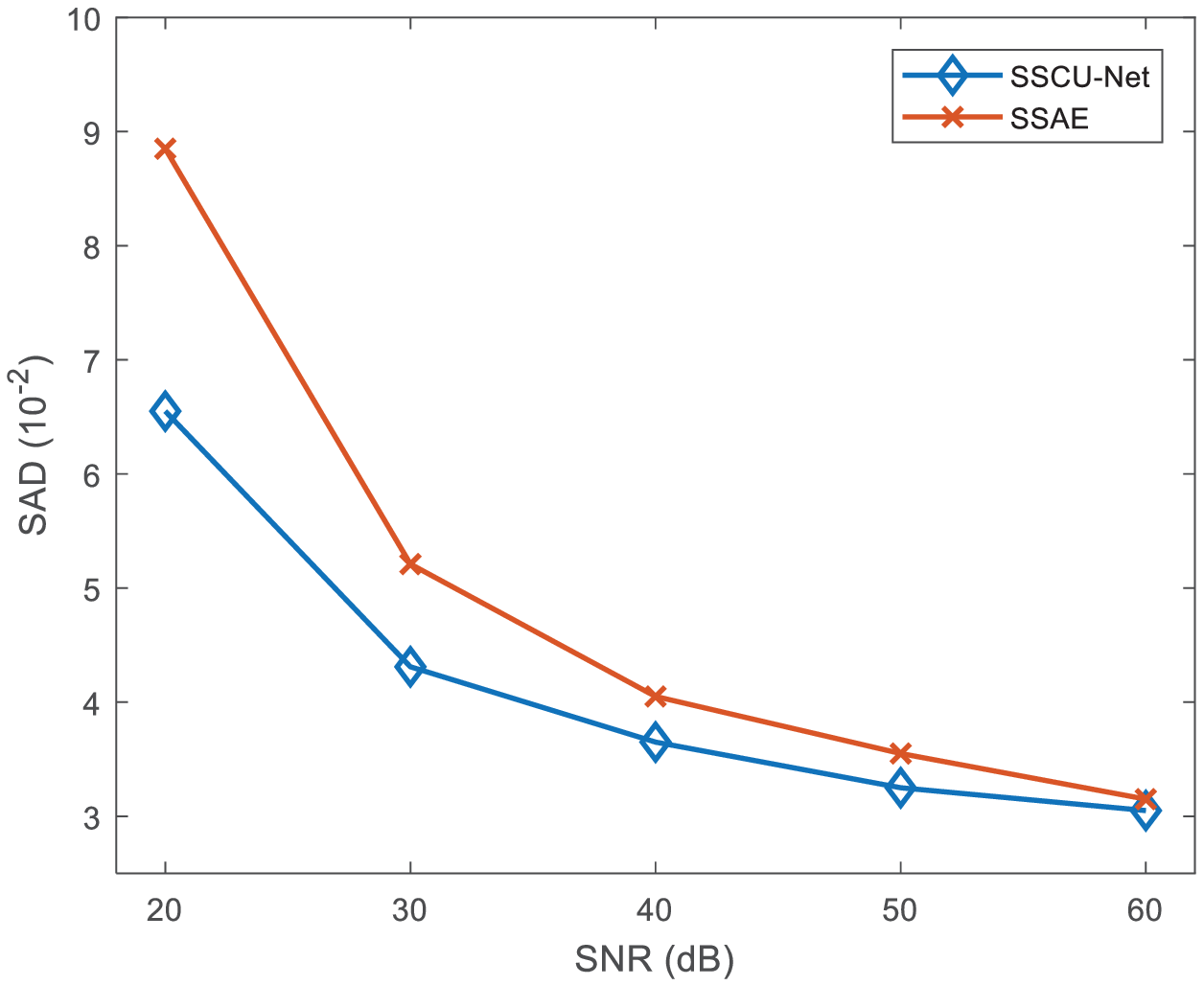}
\end{minipage}
}
\subfigure[]{
\begin{minipage}{0.46\linewidth}
\centering
\includegraphics[width=\textwidth]{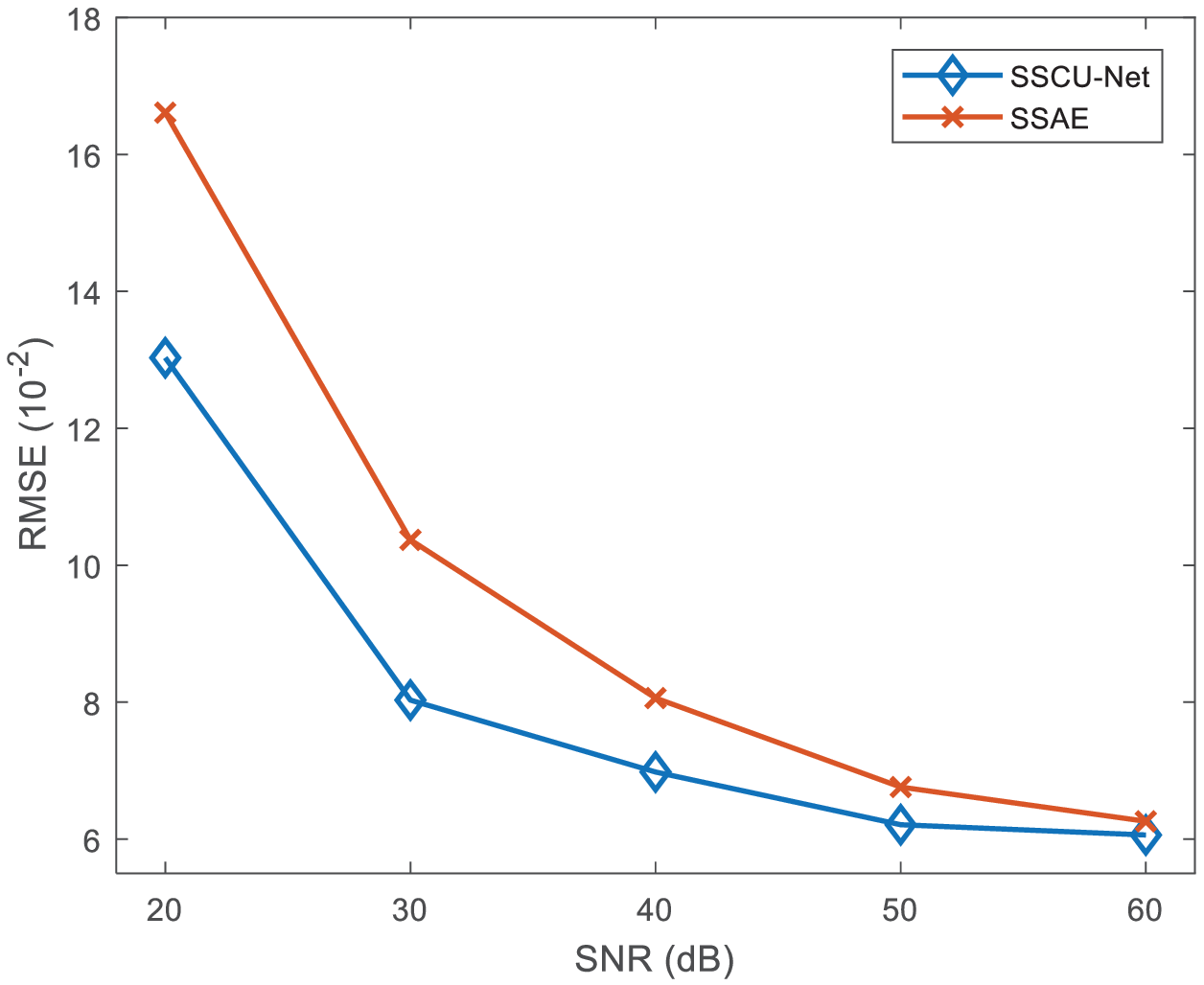}
\end{minipage}
}
\caption{Performance of the proposed SSCU-Net with respect to different noise levels. (a) Mean SAD values. (b) Mean RMSE values.}\label{noise}
\end{figure}

Meanwhile, Fig. \ref{Svsm} shows the curves of the mean SAD of endmembers under different $S$ and $m$ values on the Urban data set. It can be seen that under a fixed region size $S$, there is an optimal value of $m$. When the value of $S$ is small, the influences of $m$ on SAD are relatively limited, as shown in Fig. \ref{Svsm}(a) and (b), while the value of $S$ becomes larger, the influences of $m$ are greater, as shown in Fig. \ref{Svsm}(c)-(e). The reason for this phenomenon is that the spatial distribution of pixels in a small region is relatively uniform and the difference between pixels is small, resulting in the SAD is less affected by the shape of superpixel blocks. In a larger region, there may be a gradient change or a boundary. Under this circumstance, it makes sense to adjust the shape of superpixels to make the pixel distribution uniform.

\begin{table*}[!t]
  \centering
    \caption{SAD and RMSE ($ \times {10^{{\rm{ - }}2}}$) Unmixing Results on the Urban Data Set for the SAE Network, Mean and Standard Deviation are Computed with Running 20 Times. Best Results are in Bold.}\label{ablation1}
  \scalebox{1}{
  \tabcolsep 9pt
  \begin{tabular}{l|ccc|ccc}
  \hline \hline
\multirow{1}*{}&\multicolumn{3}{c|}{SAD (Endmember)}&\multicolumn{3}{c}{RMSE (Abundance)}\\
\multirow{1}*{Endmember}&SSAE (EENet)&SAE w/o COL&SAE w/ COL&SSAE (EENet)&SAE w/o COL&SAE w/ COL\\
\hline
\multirow{1}*{Asphalt}&6.02 $ \pm $ 0.5 &5.06 $ \pm $ 0.3 &\textbf{4.64 $ \pm $ 0.2} &15.21 $ \pm $ 2.5 &11.99 $ \pm $ 2.0 &\textbf{8.33 $ \pm $ 0.3}\\
\multirow{1}*{Grass}&3.59 $ \pm $ 0.3	&3.64 $ \pm $ 0.1	&\textbf{3.25 $ \pm $ 0.1}	&15.44 $ \pm $ 1.8 &15.43 $ \pm $ 1.6 &\textbf{8.44 $ \pm $ 0.3}\\
\multirow{1}*{Tree}&3.18 $ \pm $ 0.2 &2.93 $ \pm $ 0.2 &\textbf{2.69 $ \pm $ 0.1} &13.35 $ \pm $ 2.5 &11.81 $ \pm $ 2.1 &\textbf{5.71 $ \pm $ 0.2}\\
\multirow{1}*{Roof}&2.76 $ \pm $ 0.3 &1.80 $ \pm $ 0.2 &\textbf{1.63 $ \pm $ 0.1} &10.31 $ \pm $ 1.7 &6.62 $ \pm $ 1.1 &\textbf{5.35 $ \pm $ 0.2}\\
\multirow{1}*{Average}&3.89 $ \pm $ 0.1 &3.36 $ \pm $ 0.1 &\textbf{3.05 $ \pm $ 0.1} &13.58 $ \pm $ 0.9 &11.46 $ \pm $ 0.8 &\textbf{6.96 $ \pm $ 0.1}\\
\hline \hline
\end{tabular}}
\end{table*}

\begin{table*}[!t]
  \centering
    \caption{SAD and RMSE ($ \times {10^{{\rm{ - }}2}}$) Unmixing Results on the Urban Data Set for the SCAE Network, Mean and Standard Deviation are Computed with Running 20 Times. Best Results are in Bold.}\label{ablation2}
  \scalebox{0.83}{\begin{tabular}{l|cccc|cccc}
  \hline \hline
\multirow{1}*{}&\multicolumn{4}{c|}{SAD (Endmember)}&\multicolumn{4}{c}{RMSE (Abundance)}\\
\multirow{1}*{Endmember}&SSAE (AENet)&SSAE (AENet FW)&SCAE w/o COL&SCAE w/ COL&SSAE (AENet)&SSAE (AENet FW)&SCAE w/o COL&SCAE w/ COL\\
\hline
\multirow{1}*{Asphalt}&9.62 $ \pm $ 2.2 &6.02 $ \pm $ 0.5 &9.62 $ \pm $ 2.2 &\textbf{4.64 $ \pm $ 0.2} &11.49 $ \pm $ 2.8 &9.75 $ \pm $ 0.4 &11.49 $ \pm $ 2.8 &\textbf{8.09 $ \pm $ 0.3}\\
\multirow{1}*{Grass}&4.43 $ \pm $ 0.6	&3.59 $ \pm $ 0.3	&4.43 $ \pm $ 0.6	&\textbf{3.25 $ \pm $ 0.1} &10.59 $ \pm $ 0.9	&9.59 $ \pm $ 0.2 &10.59 $ \pm $ 0.9 &\textbf{8.28 $ \pm $ 0.2}\\
\multirow{1}*{Tree}&5.83 $ \pm $ 2.6 &3.18 $ \pm $ 0.2 &5.83 $ \pm $ 2.6 &\textbf{2.69 $ \pm $ 0.1} &8.19 $ \pm $ 1.6 &5.87 $ \pm $ 0.2 &8.19 $ \pm $ 1.6 &\textbf{5.31 $ \pm $ 0.2}\\
\multirow{1}*{Roof}&18.2 $ \pm $ 14.2 &2.76 $ \pm $ 0.3 &18.2 $ \pm $ 14.2 &\textbf{1.63 $ \pm $ 0.1} &9.42 $ \pm $ 3.3 &5.76 $ \pm $ 0.2 &9.42 $ \pm $ 3.3 &\textbf{5.02 $ \pm $ 0.2}\\
\multirow{1}*{Average}&9.52 $ \pm $ 4.5 &3.89 $ \pm $ 0.1 &9.52 $ \pm $ 4.5 &\textbf{3.05 $ \pm $ 0.1} &9.92 $ \pm $ 2.0 &7.75 $ \pm $ 0.1 &9.92 $ \pm $ 2.0 &\textbf{6.68 $ \pm $ 0.1}\\
\hline \hline
\end{tabular}}
\end{table*}

In addition, we can see that the larger the segmentation region, the smaller the optimal value of $m$, i.e., as the value of $S$ increases, the optimal $m$ point will move forward. This means that when the region becomes larger, a smaller value of $m$ is required to relax the spatial constraint between pixels. Combined with the superpixel segmentation method based on abundance information that we introduced, then the segmentation process will consider the abundance difference between pixels more to obtain a more suitable spatial shape. It can be found from Refs. \cite{qi2020spectral,huang2020spatial} that the optimal value of the window is generally $3 \times 3$, including the usage of spatial information in 3D convolution, where most of them are $3 \times 3$ spatial filters \cite{gao2021cycu}. This is mainly because the square sliding window limits the effectiveness of endmember extraction in a large region, while superpixel segmentation can further tap its potential.

\emph{2) Convergence Analysis:} The proposed SSCU-Net is alternately trained in a collaborative manner. The SAE network and SCAE network are optimized alternately, promote each other, and converge to an optimal solution together. This collaborative strategy not only makes full use of endmember and abundance information, but also accelerates the convergence speed of the networks. Fig. \ref{convergence} shows the obtained loss as a function of epochs for the SAE and SCAE networks using the synthetic data set with 30 dB SNR. Meanwhile, Fig. \ref{convergence} also shows the convergence curves of the two networks (EENet and AENet) in SSAE. The number of epochs is set to 200.

It can be seen that the convergence of SSCU-Net and SSAE is satisfactory. In the SSAE network, EENet and AENet are two independent networks. However, benefiting from the proposed collaborative strategy, the SAE network can obtain a higher quality abundance estimation vector under the collaborative influence of the SCAE network. Then it will promote the SAE network to extract a more accurate endmember matrix. It is also true for the SCAE network. In SSCU-Net, the collaborative strategy unifies the SAE network and SCAE network, which can correct and promote each other. This strategy greatly improves the convergence speed of the entire unmixing process.

\emph{3) Noise Robustness Analysis:} In order to investigate the robustness of the proposed SSCU-Net, experiments on synthetic data with different levels of noise are carried out. The SNRs are selected as 20, 30, 40, 50 and 60 dB. Fig. \ref{noise} illustrates the unmixing performance of SSCU-Net and SSAE regarding SNR settings. As expected, with the increasing in SNR, both SSCU-Net and SSAE show a decreasing trend in SAD and RMSE. Meanwhile, we can observe that SSCU-Net shows better SAD and RMSE values than SSAE under all SNR settings, especially on low SNR data sets. This is mainly due to its inherent advantages of more effective expression of the spatial-spectral structure of HSIs. It also attributes to the adopted collaborative learning strategy, where SSCU-Net is equipped with the hybrid advantages of SAE based on spatial information and SCAE based on spectral information.

\begin{table*}[!t]
  \centering
    \caption{Endmember Extraction Results ($ \times {10^{{\rm{ - }}2}}$) for the Urban Data Set, Mean and Standard Deviation are Computed with Running 20 Times. Best Results are in Bold.}\label{SADub}
  \scalebox{1}{
  \tabcolsep 6pt
  \begin{tabular}{l|cccccccc}
  \hline \hline
\multirow{1}*{Endmember (SAD)}&VCA&${L_{1/2}}$-NMF&Dgs-NMF&EndNet&TANet&CNNAEU&SSAE&SSCU-Net\\
\hline
\multirow{1}*{Asphalt}&18.85 $ \pm $ 7.2 &6.06 $ \pm $ 0.2 &5.86 $ \pm $ 0.1 &5.98 $ \pm $ 0.2 &\textbf{3.86 $ \pm $ 0.3} &5.75 $ \pm $ 0.5 &6.02 $ \pm $ 0.5 &4.64 $ \pm $ 0.2\\
\multirow{1}*{Grass}&33.48 $ \pm $ 13.0 &20.05 $ \pm $ 0.4 &13.69 $ \pm $ 0.4 &5.34 $ \pm $ 0.1 &3.66 $ \pm $ 0.3 &3.66 $ \pm $ 0.4 &3.59 $ \pm $ 0.3 &\textbf{3.25 $ \pm $ 0.1}\\
\multirow{1}*{Tree}&18.35 $ \pm $ 13.0 &3.71 $ \pm $ 0.0 &4.12 $ \pm $ 0.0 &4.57 $ \pm $ 0.2 &7.49 $ \pm $ 0.4 &3.21 $ \pm $ 0.3 &3.18 $ \pm $ 0.2 &\textbf{2.69 $ \pm $ 0.1}\\
\multirow{1}*{Roof}&71.42 $ \pm $ 0.2 &14.22 $ \pm $ 0.2 &10.54 $ \pm $ 0.3 &3.89 $ \pm $ 0.7 &2.51 $ \pm $ 0.3 &3.32 $ \pm $ 0.6 &2.76 $ \pm $ 0.3 &\textbf{1.63 $ \pm $ 0.1}\\
\hline
\multirow{1}*{Average}&35.52 $ \pm $ 4.1 &11.01 $ \pm $ 0.2 &8.85 $ \pm $ 0.2 &4.95 $ \pm $ 0.3 &4.38 $ \pm $ 0.3 &3.98 $ \pm $ 0.3 &3.89 $ \pm $ 0.1 &\textbf{3.05 $ \pm $ 0.1}\\
\hline \hline
\end{tabular}}
\end{table*}

\begin{table*}[!t]
  \centering
    \caption{Abundance Estimation Results ($ \times {10^{{\rm{ - }}2}}$) for the Urban Data Set, Mean and Standard Deviation are Computed with Running 20 Times. Best Results are in Bold.}\label{RMSEub}
  \scalebox{1}{
  \tabcolsep 6pt
  \begin{tabular}{l|cccccccc}
  \hline \hline
\multirow{1}*{Abundance (RMSE)}&VCA&${L_{1/2}}$-NMF&Dgs-NMF&EndNet&TANet&CNNAEU&SSAE&SSCU-Net\\
\hline
\multirow{1}*{Asphalt}&33.05 $ \pm $ 19.7 &14.77 $ \pm $ 0.1 &13.18 $ \pm $ 0.1 &10.62 $ \pm $ 0.1 &\textbf{7.03 $ \pm $ 0.3} &12.49 $ \pm $ 0.3 &9.75 $ \pm $ 0.4 &8.09 $ \pm $ 0.3\\
\multirow{1}*{Grass}&46.56 $ \pm $ 7.9 &16.16 $ \pm $ 0.2 &12.95 $ \pm $ 0.0 &13.85 $ \pm $ 0.3 &12.07 $ \pm $ 0.4 &12.57 $ \pm $ 0.3 &9.59 $ \pm $ 0.2 &\textbf{8.28 $ \pm $ 0.2}\\
\multirow{1}*{Tree}&31.32 $ \pm $ 20.2 &12.65 $ \pm $ 0.2 &9.57 $ \pm $ 0.1 &9.07 $ \pm $ 0.3 &6.37 $ \pm $ 0.4 &8.55 $ \pm $ 0.2 &5.87 $ \pm $ 0.2 &\textbf{5.31 $ \pm $ 0.2}\\
\multirow{1}*{Roof}&17.25 $ \pm $ 3.2 &6.90 $ \pm $ 0.1 &6.27 $ \pm $ 0.0 &6.51 $ \pm $ 0.2 &9.77 $ \pm $ 0.3 &8.54 $ \pm $ 0.2 &5.76 $ \pm $ 0.2 &\textbf{5.02 $ \pm $ 0.2}\\
\hline
\multirow{1}*{Average}&32.04 $ \pm $ 2.4 &12.62 $ \pm $ 0.1 &10.49 $ \pm $ 0.1 &10.02 $ \pm $ 0.2 &8.81 $ \pm $ 0.3 &10.54 $ \pm $ 0.1 &7.75 $ \pm $ 0.1 &\textbf{6.68 $ \pm $ 0.1}\\
\hline \hline
\end{tabular}}
\end{table*}

\begin{figure*}[t]
\begin{center}
\includegraphics[width=1\linewidth]{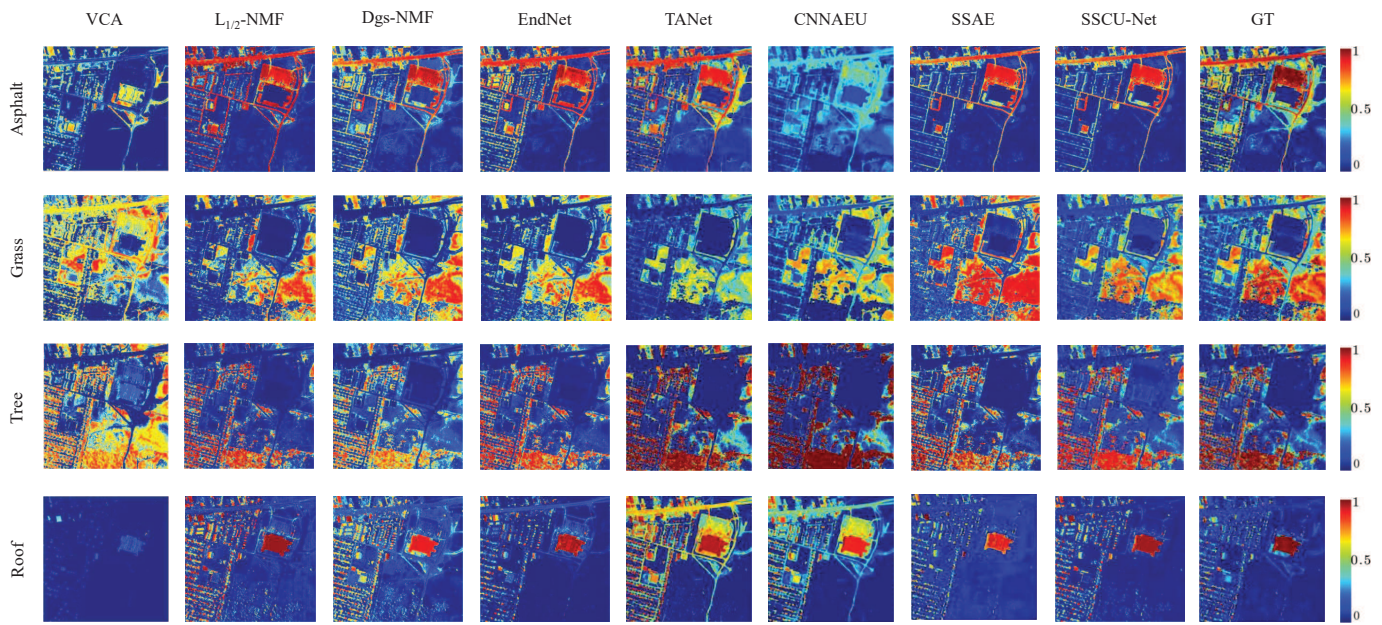}
\end{center}
   \caption{Visual abundance maps under different methods and the GTs on the Urban data set.}
\label{abuub}
\end{figure*}

\begin{figure*}[!t]
\centering
\subfigure[]{
\begin{minipage}{0.23\linewidth}
\centering
\includegraphics[width=\textwidth]{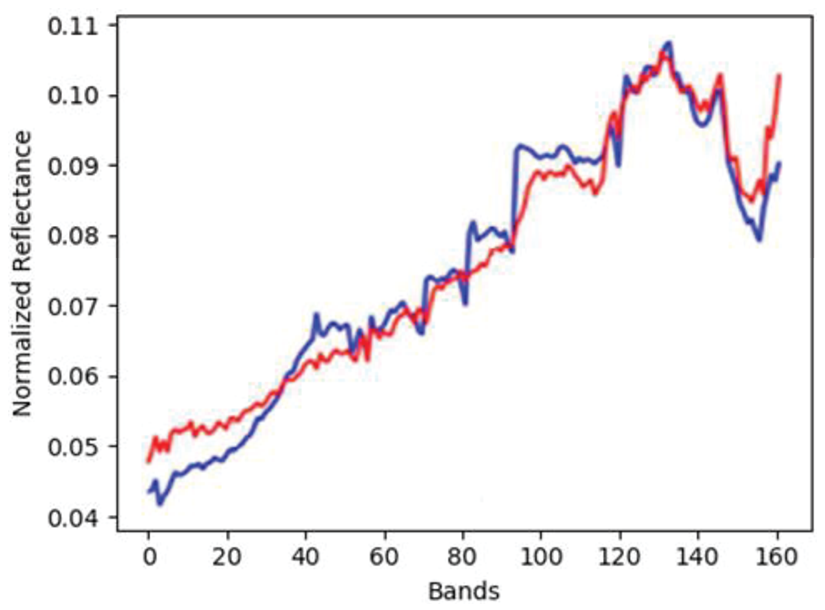}
\end{minipage}
}
\subfigure[]{
\begin{minipage}{0.23\linewidth}
\centering
\includegraphics[width=\textwidth]{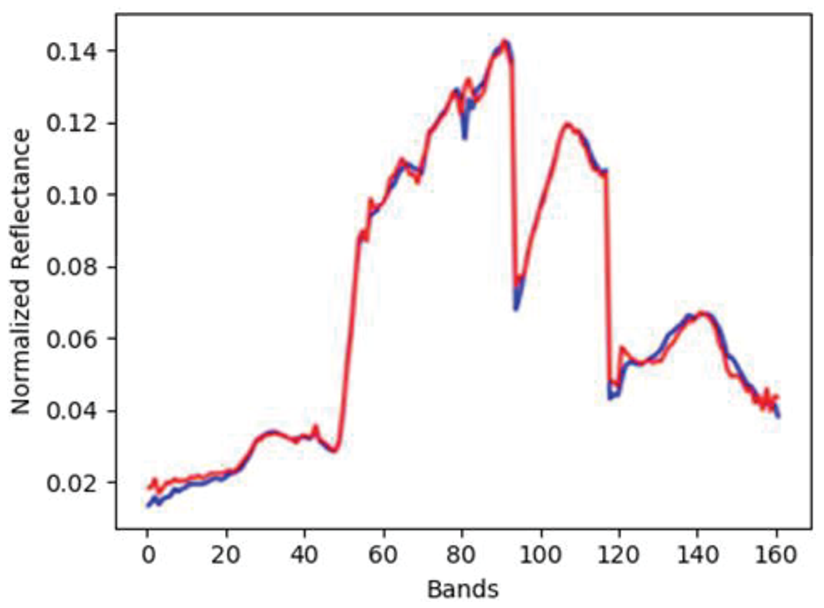}
\end{minipage}
}
\subfigure[]{
\begin{minipage}{0.23\linewidth}
\centering
\includegraphics[width=\textwidth]{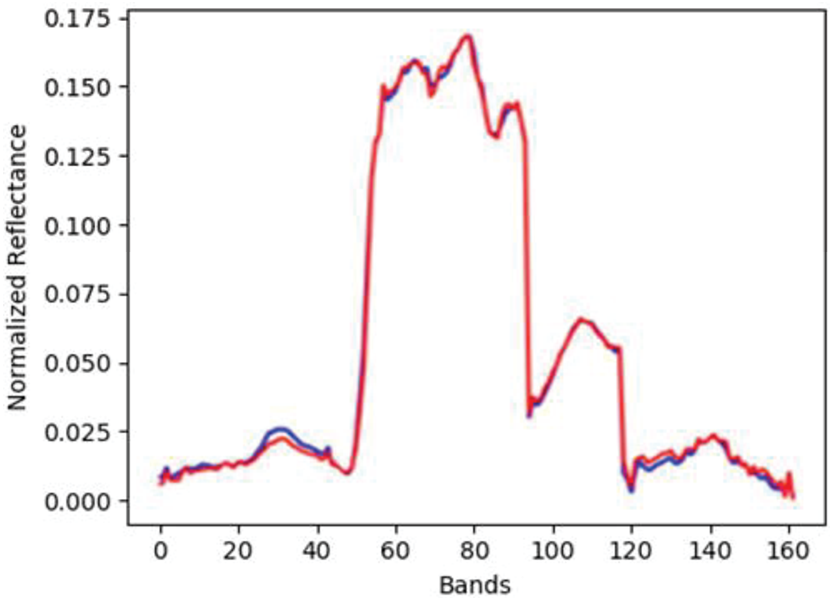}
\end{minipage}
}
\subfigure[]{
\begin{minipage}{0.23\linewidth}
\centering
\includegraphics[width=\textwidth]{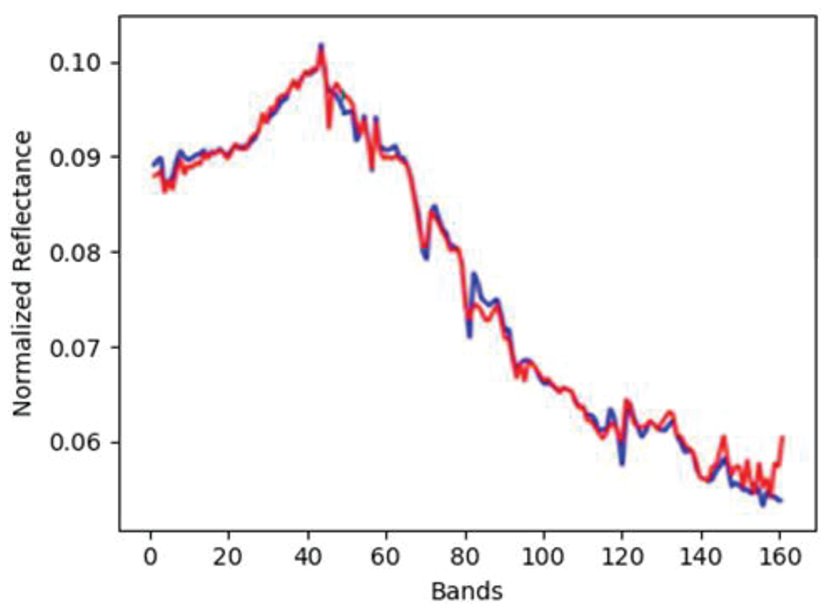}
\end{minipage}
}
\caption{Endmember comparison between SSCU-Net (blue curves) and the corresponding GT (red curves) on the Urban data set. (a) Endmember Asphalt. (b) Endmember Grass. (c) Endmember Tree. (d) Endmember Roof.}\label{EMub}
\end{figure*}

\subsection{Ablation Study}
In the two-stream SSCU-Net, both the SAE network and SCAE network can be used as independent unmixing networks, and they build an efficient unmixing network via collaborative learning strategy.
We explore the improvements benefiting from the two key concepts in the proposed framework: superpixel-based spatial information processing and collaborative learning.
Meanwhile, we discuss some discoveries on the application of spatial information and spectral information in AE-based unmixing networks.

\emph{1) Experiments on SAE Network:}
In the SAE network, spatial information based on superpixel segmentation is introduced. Different from the spatial information used in SSAE, it can extract endmembers more effectively.
Table \ref{ablation1} shows the unmixing results of SAE network on the Urban data set, and the unmixing results of EENet in SSAE. It should be pointed out that the SSAE network consists of two independent networks, EENet and AENet. EENet is used to extract endmembers, while AENet is used for abundance estimation. The decoder weight of AENet is not trainable, and it is fixed as the endmembers extracted by EENet. Table \ref{ablation1} shows the unmixing results of EENet as an independent unmixing network. We can see that the independent SAE network based on superpixel segmentation (SAE w/o COL) has an improvement in both endmember extraction and abundance estimation compared with the sliding window-based EENet. After using the collaborative strategy, the SAE network (SAE w/ COL) is further improved. Under the influence of the collaborative strategy, the abundances are greatly improved due to the positive impact of SCAE network.

\begin{table*}[!t]
  \centering
    \caption{Endmember Extraction Results ($ \times {10^{{\rm{ - }}2}}$) for the Jasper Ridge Data Set, Mean and Standard Deviation are Computed with Running 20 Times. Best Results are in Bold.}\label{SADjr}
  \scalebox{1}{
  \tabcolsep 6pt
  \begin{tabular}{l|cccccccc}
  \hline \hline
\multirow{1}*{Endmember (SAD)}&VCA&${L_{1/2}}$-NMF&Dgs-NMF&EndNet&TANet&CNNAEU&SSAE&SSCU-Net\\
\hline
\multirow{1}*{Tree}&13.95 $ \pm $ 3.1 &15.10 $ \pm $ 0.3 &4.66 $ \pm $ 0.2 &4.57 $ \pm $ 0.4 &2.83 $ \pm $ 0.2 &11.94 $ \pm $ 2.1 &3.37 $ \pm $ 0.1 &\textbf{2.57 $ \pm $ 0.1}\\
\multirow{1}*{Water}&29.09 $ \pm $ 10.2 &4.60 $ \pm $ 0.0 &4.60 $ \pm $ 0.0 &5.05 $ \pm $ 0.9 &\textbf{3.31 $ \pm $ 0.3} &6.92 $ \pm $ 0.4 &4.72 $ \pm $ 0.1 &3.74 $ \pm $ 0.1\\
\multirow{1}*{Soil}&15.22 $ \pm $ 2.3 &6.16 $ \pm $ 0.5 &5.66 $ \pm $ 0.2 &5.29 $ \pm $ 0.3 &8.48 $ \pm $ 0.4 &10.15 $ \pm $ 0.9 &3.02 $ \pm $ 0.3 &\textbf{2.04 $ \pm $ 0.3}\\
\multirow{1}*{Road}&9.94 $ \pm $ 1.1 &9.81 $ \pm $ 0.1 &6.73 $ \pm $ 0.1 &3.54 $ \pm $ 0.2 &2.14 $ \pm $ 0.2 &7.45 $ \pm $ 0.4 &2.77 $ \pm $ 0.2 &\textbf{1.95 $ \pm $ 0.2}\\
\hline
\multirow{1}*{Average}&17.05 $ \pm $ 3.6 &7.19 $ \pm $ 2.4 &5.41 $ \pm $ 0.1 &4.61 $ \pm $ 0.5 &4.19 $ \pm $ 0.3 &9.12 $ \pm $ 0.6 &3.47 $ \pm $ 0.1 &\textbf{2.58 $ \pm $ 0.1}\\
\hline \hline
\end{tabular}}
\end{table*}

\begin{table*}[!t]
  \centering
    \caption{Abundance Estimation Results ($ \times {10^{{\rm{ - }}2}}$) for the Jasper Ridge Data Set, Mean and Standard Deviation are Computed with Running 20 Times. Best Results are in Bold.}\label{RMSEjr}
  \scalebox{1}{
  \tabcolsep 6pt
  \begin{tabular}{l|cccccccc}
  \hline \hline
\multirow{1}*{Abundance (RMSE)}&VCA&${L_{1/2}}$-NMF&Dgs-NMF&EndNet&TANet&CNNAEU&SSAE&SSCU-Net\\
\hline
\multirow{1}*{Tree}&16.09 $ \pm $ 3.1 &16.16 $ \pm $ 0.5 &11.66 $ \pm $ 0.2 &8.85 $ \pm $ 0.4 &8.01 $ \pm $ 0.4 &13.56 $ \pm $ 0.3	&5.15 $ \pm $ 0.6 &\textbf{4.21 $ \pm $ 0.6}\\
\multirow{1}*{Water}&6.06 $ \pm $ 1.8 &5.57 $ \pm $ 0.0 &\textbf{4.13 $ \pm $ 0.0} &6.88 $ \pm $ 0.3 &8.12 $ \pm $ 0.3 &9.66 $ \pm $ 0.3 &5.63 $ \pm $ 0.4 &4.66 $ \pm $ 0.4\\
\multirow{1}*{Soil}&15.00 $ \pm $ 1.5 &17.02 $ \pm $ 0.4 &11.13 $ \pm $ 0.3 &10.59 $ \pm $ 0.2 &8.59 $ \pm $ 0.4 &10.61 $ \pm $ 0.2	&6.19 $ \pm $ 0.4 &\textbf{5.38 $ \pm $ 0.4}\\
\multirow{1}*{Road}&11.09 $ \pm $ 1.5 &6.73 $ \pm $ 0.2 &5.68 $ \pm $ 0.1 &11.17 $ \pm $ 0.4 &\textbf{5.45 $ \pm $ 0.3} &8.64 $ \pm $ 0.2 &7.15 $ \pm $ 0.3 &6.23 $ \pm $ 0.3\\
\hline
\multirow{1}*{Average}&12.05 $ \pm $ 1.5 &11.37 $ \pm $ 0.2 &8.15 $ \pm $ 0.2 &9.37 $ \pm $ 0.5 &7.54 $ \pm $ 0.3 &10.62 $ \pm $ 0.1 &6.03 $ \pm $ 0.2 &\textbf{5.12 $ \pm $ 0.2}\\
\hline \hline
\end{tabular}}
\end{table*}

\emph{2) Experiments on SCAE Network:}
In the SCAE network, as in SSAE, we use a 1D convolutional network to extract effective spectral characteristics to obtain better abundance estimation results. Table \ref{ablation2} illustrates the unmixing results of SCAE network on the Urban data set, and the results of AENet in SSAE. It is worth noting that in SSAE, the endmembers of AENet are fixed and not learnable. This strategy is also used in CNNAEU to estimate abundances. In order to deeply explore the role of spectral convolutional network in AE-based unmixing network, we respectively give the unmixing results of AENet as an independent unmixing network and when its endmembers are fixed. The result in the case of fixed weights is expressed as AENet FW. It should be also noted that when the SCAE is used as an independent unmixing network, i.e., SCAE w/o COL, its structure is the same as AENet.

From Table \ref{ablation2}, we can observe that the unmixing result of the independent SCAE network is the same as that of AENet. When the endmembers are fixed, that is, the decoder weight of AENet is fixed to the endmember matrix extracted by EENet, the abundances estimated by AENet will be further improved. After the introduction of the collaborative strategy, the endmembers extracted by the SCAE network (SCAE w/ COL) are the same as those of the SAE network, since they share the decoder weight with each other. Under the influence of the collaborative strategy, better endmembers promote the best result of the abundances estimated by the SCAE network.

\begin{figure*}[t]
\begin{center}
\includegraphics[width=1\linewidth]{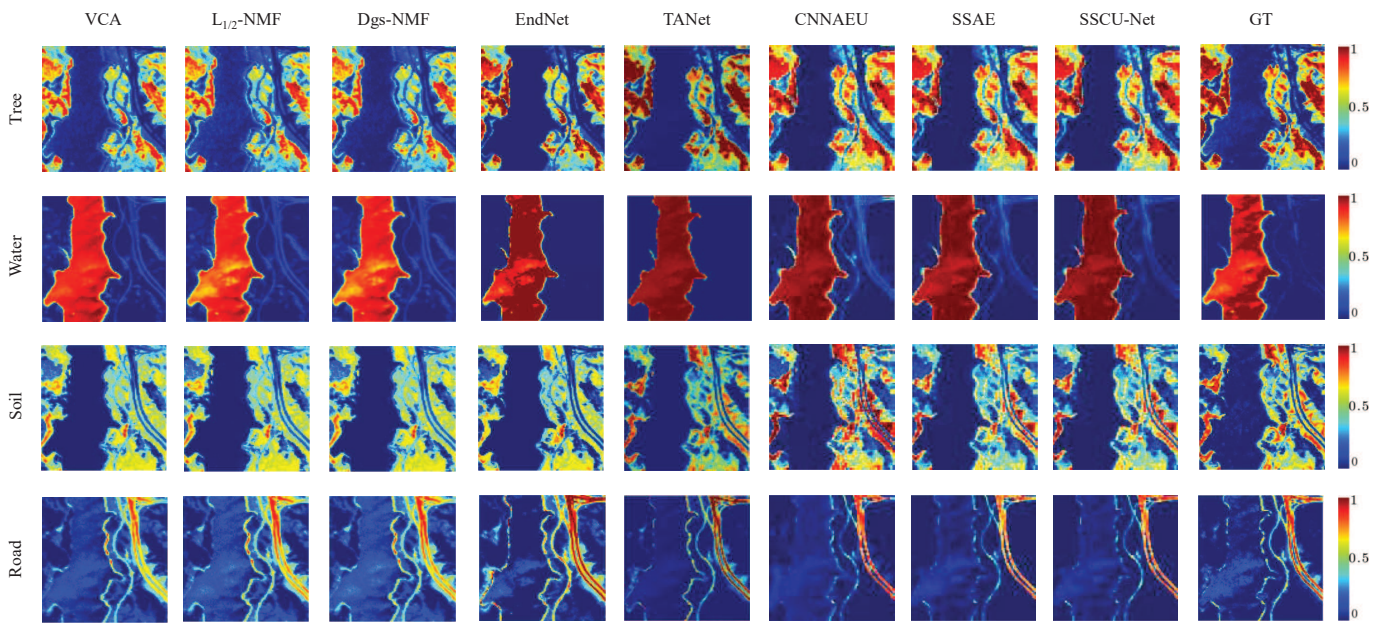}
\end{center}
   \caption{Visual abundance maps under different methods and the GTs on the Jasper Ridge data set.}
\label{abujr}
\end{figure*}

\begin{figure*}[!t]
\centering
\subfigure[]{
\begin{minipage}{0.23\linewidth}
\centering
\includegraphics[width=\textwidth]{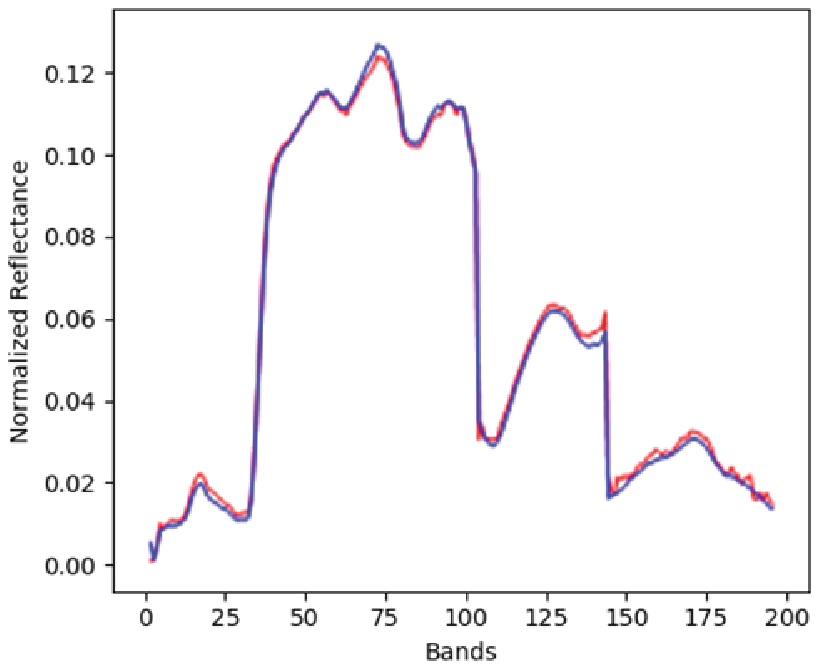}
\end{minipage}
}
\subfigure[]{
\begin{minipage}{0.23\linewidth}
\centering
\includegraphics[width=\textwidth]{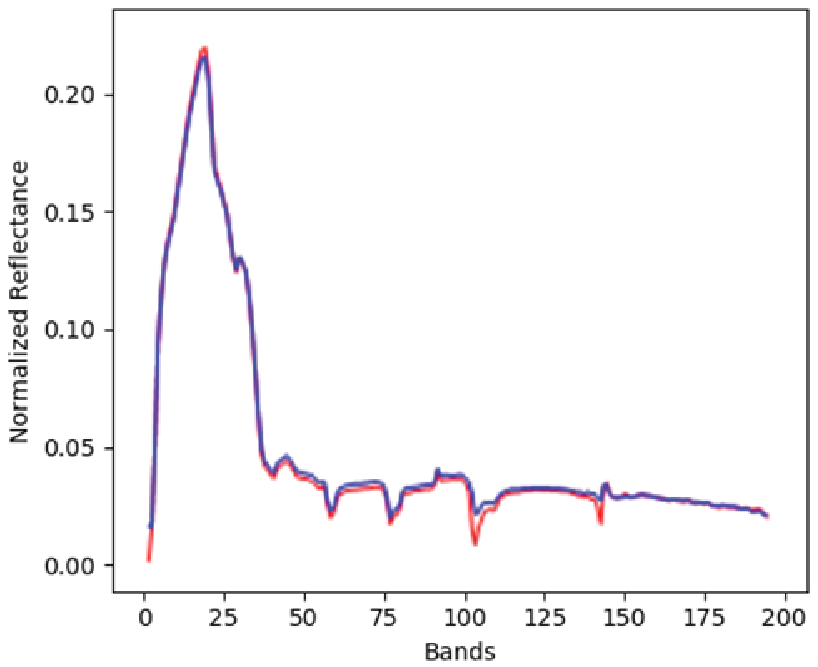}
\end{minipage}
}
\subfigure[]{
\begin{minipage}{0.23\linewidth}
\centering
\includegraphics[width=\textwidth]{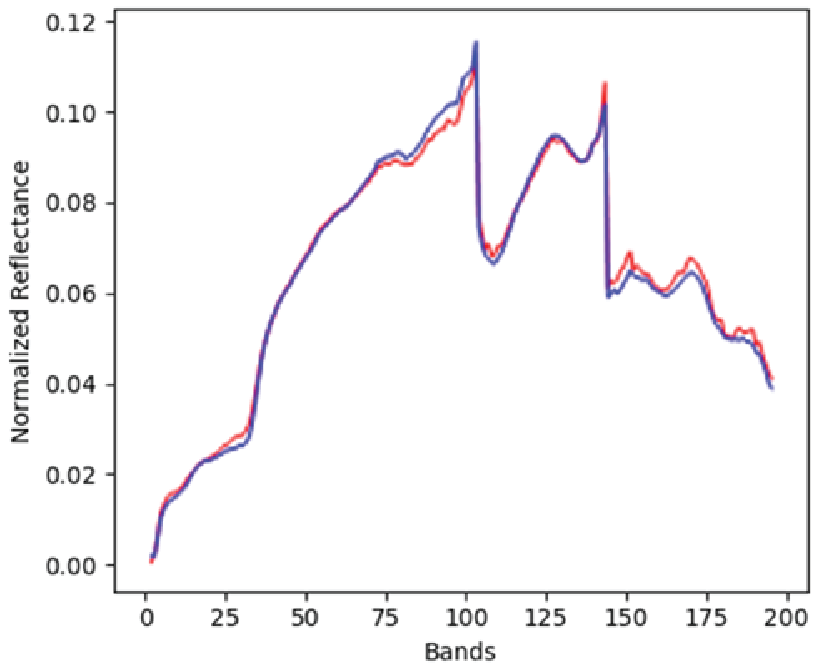}
\end{minipage}
}
\subfigure[]{
\begin{minipage}{0.23\linewidth}
\centering
\includegraphics[width=\textwidth]{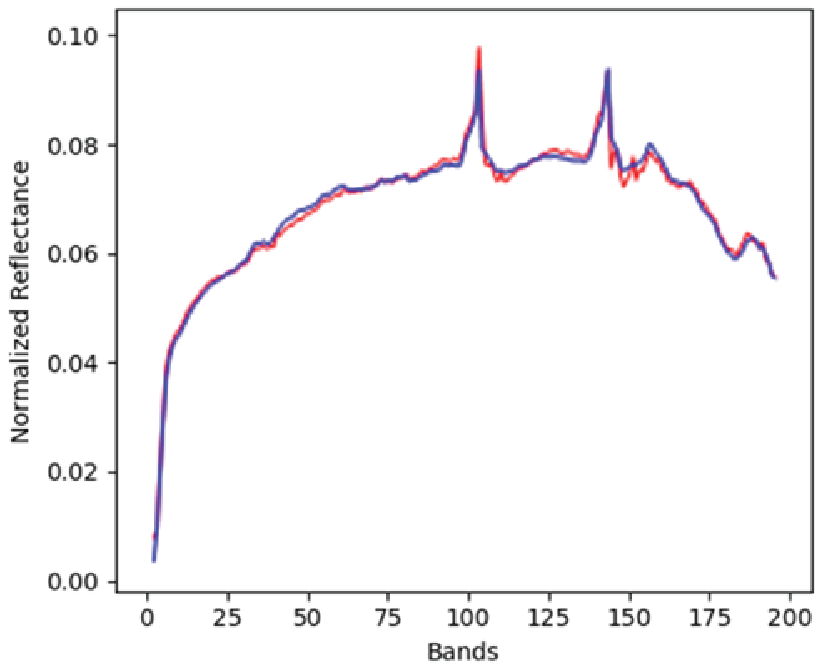}
\end{minipage}
}
\caption{Endmember comparison between SSCU-Net (blue curves) and the corresponding GT (red curves) on the Jasper Ridge data set. (a) Endmember Tree. (b) Endmember Water. (c) Endmember Soil. (d) Endmember Road.}\label{EMjr}
\end{figure*}

\emph{3) Discussion:}
By comparing Table \ref{ablation1} and Table \ref{ablation2}, we can find that the AE-based unmixing network that explores spatial information, whether EENet or SAE w/o COL, has a very good endmember extraction effect, but its abundance estimation result is unsatisfactory. In other words, AE networks that introduce spatial information tend to obtain better endmembers. On the other hand, the convolutional AE-based unmixing network that explores spectral information, including AENet and SCAE w/o COL, has poorly extracted endmembers. However, compared with EENet and SAE w/o COL, in the case of worse endmembers, the abundance obtained is better. In other words, the convolutional AE network using spectral information is more inclined to obtain better abundances. In the case of fixing the endmembers of AENet, the abundances estimated by AENet FW has been further improved. These experimental results strongly support the statement in Section III.

The collaborative strategy between endmembers and abundances is proposed to give full play to the respective advantages of the two-stream network. As shown in the experimental results, SAE w/ COL and SCAE w/ COL, i.e., SSCU-Net, achieve the best unmixing results, both in terms of endmembers and abundances. It should be pointed out that SSCU-Net eventually obtains an endmember matrix and two abundance matrices, which are determined by the proposed collaborative strategy. We can see that the final abundance obtained by the SCAE network is better than that of SAE. Therefore, the abundance estimated by the SCAE network is selected as the final abundance of SSCU-Net.

\subsection{Comparison of SSCU-Net Against Other Methods}
\emph{1) Experiments on Urban Data Set:}
Table \ref{SADub} shows the endmember extraction results of all comparison algorithms on the Urban data set, including the SAD value for each endmember. We can see that benefiting from the powerful data fitting capabilities of deep learning, AE-based unmixing networks have generally achieved better endmember extraction results than traditional NMF-based methods. Due to the consideration of the endmember guidance, TANet obtains better endmembers compared with EndNet. Spatial-spectral-based methods, including CNNAEU, SSAE, and SSCU-Net, have more advantages than the spectral-based methods. The endmembers extracted by SSAE and CNNAEU are similar, and they have more than 1${\rm{\% }}$ improvement over the endmembers extracted by EndNet. As expected, the two-stream network SSCU-Net that efficiently utilizes spatial-spectral information can extract the best endmembers.

Table \ref{RMSEub} details the abundance estimation results of all unmixing methods. The results show the same trend as the endmember extraction results. The difference is that EndNet and CNNAEU have no obvious advantages compared to traditional methods. The abundances estimated by TANet have improved, and it obtains the best result on endmember Asphalt. SSAE and SSCU-Net, which employ convolutional AE networks, have achieved 3${\rm{\% }}$ and 4${\rm{\% }}$ abundance improvement respectively. This also confirms the advantages of the convolutional network in abundance estimation and the excellence of the two-stream network SSCU-Net.

For illustrative purposes, Fig. \ref{abuub} visualizes the abundance maps acquired by all methods, corresponding to the estimation results in Table \ref{RMSEub}. Compared with other unmixing networks based on AE, SSCU-Net shows better abundance details and is closer to the ground truth (GT). Meanwhile, Fig. \ref{EMub} shows the visual comparison of endmembers between our SSCU-Net and GTs. It can be seen that SSCU-Net obtains good endmember estimations, in terms of shapes and band absorption.

\begin{table*}[!t]
  \centering
    \caption{Endmember Extraction Results ($ \times {10^{{\rm{ - }}2}}$) for the Samson Data Set, Mean and Standard Deviation are Computed with Running 20 Times. Best Results are in Bold.}\label{SADsam}
  \scalebox{1}{
  \tabcolsep 6pt
  \begin{tabular}{l|cccccccc}
  \hline \hline
\multirow{1}*{Endmember (SAD)}&VCA&${L_{1/2}}$-NMF&Dgs-NMF&EndNet&TANet&CNNAEU&SSAE&SSCU-Net\\
\hline
\multirow{1}*{Soil}&4.21 $ \pm $ 0.0 &6.21 $ \pm $ 7.3 &5.64 $ \pm $ 7.4 &1.98 $ \pm $ 0.2 &\textbf{1.22 $ \pm $ 0.2} &3.73 $ \pm $ 2.1 &2.04 $ \pm $ 0.1 &1.38 $ \pm $ 0.1\\
\multirow{1}*{Tree}&5.58 $ \pm $ 0.2 &5.23 $ \pm $ 0.3 &4.80 $ \pm $ 0.3 &5.31 $ \pm $ 0.3 &3.31 $ \pm $ 0.3 &3.97 $ \pm $ 0.4 &3.61 $ \pm $ 0.1 &\textbf{2.70 $ \pm $ 0.1}\\
\multirow{1}*{Water}&43.97 $ \pm $ 31.1 &11.97 $ \pm $ 2.1 &4.70 $ \pm $ 0.3 &3.98 $ \pm $ 0.2 &6.22 $ \pm $ 0.3 &4.30 $ \pm $ 0.9 &3.12 $ \pm $ 0.2 &\textbf{2.57 $ \pm $ 0.2}\\
\hline
\multirow{1}*{Average}&17.92 $ \pm $ 1.5 &7.80 $ \pm $ 3.2 &5.05 $ \pm $ 2.7 &3.76 $ \pm $ 0.2 &3.58 $ \pm $ 0.3 &4.00 $ \pm $ 0.7 &2.92 $ \pm $ 0.1 &\textbf{2.22 $ \pm $ 0.1}\\
\hline \hline
\end{tabular}}
\end{table*}

\begin{table*}[!t]
  \centering
    \caption{Abundance Estimation Results ($ \times {10^{{\rm{ - }}2}}$) for the Samson Data Set, Mean and Standard Deviation are Computed with Running 20 Times. Best Results are in Bold.}\label{RMSEsam}
  \scalebox{1}{
  \tabcolsep 6pt
  \begin{tabular}{l|cccccccc}
  \hline \hline
\multirow{1}*{Abundance (RMSE)}&VCA&${L_{1/2}}$-NMF&Dgs-NMF&EndNet&TANet&CNNAEU&SSAE&SSCU-Net\\
\hline
\multirow{1}*{Soil}&16.45 $ \pm $ 3.0 &8.58 $ \pm $ 3.3 &7.77 $ \pm $ 3.8 &8.60 $ \pm $ 0.0 &6.96 $ \pm $ 0.3 &18.49 $ \pm $ 0.6 &4.70 $ \pm $ 0.4 &\textbf{4.04 $ \pm $ 0.3}\\
\multirow{1}*{Tree}&11.25 $ \pm $ 1.0 &7.44 $ \pm $ 3.7 &7.74 $ \pm $ 3.6 &6.92 $ \pm $ 0.1 &5.71 $ \pm $ 0.3 &16.21 $ \pm $ 0.4 &5.33 $ \pm $ 0.3 &\textbf{4.35 $ \pm $ 0.3}\\
\multirow{1}*{Water}&19.21 $ \pm $ 3.3 &5.55 $ \pm $ 0.9 &\textbf{2.70 $ \pm $ 0.9} &4.99 $ \pm $ 0.0 &4.90 $ \pm $ 0.3 &6.69 $ \pm $ 0.3 &5.27 $ \pm $ 0.3 &4.78 $ \pm $ 0.3\\
\hline
\multirow{1}*{Average}&15.64 $ \pm $ 0.6 &7.19 $ \pm $ 2.4 &6.07 $ \pm $ 2.8 &6.84 $ \pm $ 0.0 &5.86 $ \pm $ 0.3 &13.80 $ \pm $ 0.4 &5.10 $ \pm $ 0.2 &\textbf{4.39 $ \pm $ 0.2}\\
\hline \hline
\end{tabular}}
\end{table*}

\emph{2) Experiments on Jasper Ridge Data Set:}
Table \ref{SADjr} shows the endmember extraction results of all comparison algorithms on the Jasper Ridge data set. Similarly, the unmixing methods based on deep learning achieve better endmember estimation results. However, CNNAEU does not perform well on this data set. One possible reason is related to the size of Jasper Ridge data and 3D convolution. The role of 3D convolution may not be fully utilized on a small data set. The proposed SSCU-Net using superpixel segmentation can adaptively adjust the spatial shape and size to extract higher quality endmembers.

Table \ref{RMSEjr} details the abundance estimation results of all unmixing methods. Compared with traditional methods or TANet, CNNAEU and EndNet have not improved significantly. The abundance estimation result of CNNAEU is unsatisfactory among the unmixing methods based on AE. We notice that CNNAEU shows high variance in the process of abundance estimation \cite{palsson2021convolutional}. This also implies that the application of 3D convolutional networks in unmixing still needs to be further explored. SSAE and SSCU-Net, which use 1D convolutional networks based on spectral information, show their effectiveness and generalization capabilities.

Fig. \ref{abujr} visualizes the abundance maps obtained by all methods. Echoing the results in Table \ref{RMSEjr}, the abundance maps of SSAE and SSCU-Net perform the best and are also very close to GTs. Similarly, Fig. \ref{EMjr} shows a visual comparison of endmembers between SSCU-Net and GTs. Compared with the endmember estimation on the Urban data set, the endmembers estimated by SSCU-Net on the Jasper Ridge data set are closer to GTs, since the Urban data set is more complex.

\emph{3) Experiments on Samson Data Set:}
Tables \ref{SADsam} and \ref{RMSEsam} show the endmember extraction and abundance estimation results of all comparison methods on the Samson data set. The overall performances are similar to that on the Jasper Ridge data set. In terms of endmember extraction, all the AE-based unmixing networks have achieved satisfactory results, including CNNAEU. Unsurprisingly, SSCU-Net has greater advantages, except for individual endmember.

In the results of abundance estimation, EndNet and CNNAEU still have no advantages. TANet, SSAE and SSCU-Net have obtained very good abundances. SSCU-Net still achieves the best results, showing its effectiveness and good stability.

\section{Conclusion}
In this article, we have developed a new spatial-spectral collaborative unmixing network architecture to make full use of the rich spectral and spatial information in HSI. Based on some important findings on the different roles of spatial and spectral information in AE-based unmixing networks, we propose a two-stream network architecture, which consists of a spatial AE network based on superpixel segmentation and a spectral AE network based on convolutional network. It shares an alternating architecture, and can be efficiently trained in a collaborative way of endmembers and abundances. Experimental results on both synthetic and real hyperspectral data demonstrate the excellent performance of the proposed SSCU-Net architecture. Despite the achieved encouraging experimental results in this article, there are still some issues worthy of future investigation. SSCU-Net is a collaborative unmixing framework that efficiently utilizes spatial-spectral information. The two AE networks of SAE and SCAE in the SSCU-Net have adopted a simple network structure, and a deeper and more effective network structure needs to be designed to further enhance the unmixing effect. In addition, more advanced reconstruction loss functions (SAD used in SSCU-Net) are worth exploring to further optimize the network architecture.

\bibliography{sources}

\begin{thebibliography}{10}
\providecommand{\url}[1]{#1}
\csname url@samestyle\endcsname
\providecommand{\newblock}{\relax}
\providecommand{\bibinfo}[2]{#2}
\providecommand{\BIBentrySTDinterwordspacing}{\spaceskip=0pt\relax}
\providecommand{\BIBentryALTinterwordstretchfactor}{4}
\providecommand{\BIBentryALTinterwordspacing}{\spaceskip=\fontdimen2\font plus
\BIBentryALTinterwordstretchfactor\fontdimen3\font minus
  \fontdimen4\font\relax}
\providecommand{\BIBforeignlanguage}[2]{{%
\expandafter\ifx\csname l@#1\endcsname\relax
\typeout{** WARNING: IEEEtran.bst: No hyphenation pattern has been}%
\typeout{** loaded for the language `#1'. Using the pattern for}%
\typeout{** the default language instead.}%
\else
\language=\csname l@#1\endcsname
\fi
#2}}
\providecommand{\BIBdecl}{\relax}
\BIBdecl

\bibitem{shippert2004use}
P.~Shippert, ``Why use hyperspectral imagery?'' \emph{Photogramm. Eng. Remote
  Sens.}, vol.~70, no.~4, pp. 377--396, Apr. 2004.

\bibitem{plaza2011foreword}
A.~Plaza, Q.~Du, J.~M. Bioucas-Dias, X.~Jia, and F.~A. Kruse, ``Foreword to the
  special issue on spectral unmixing of remotely sensed data,'' \emph{IEEE
  Trans. Geosci. Remote Sens.}, vol.~49, no.~11, pp. 4103--4110, Nov. 2011.

\bibitem{keshava2002spectral}
N.~Keshava and J.~F. Mustard, ``Spectral unmixing,'' \emph{IEEE Signal Process.
  Mag.}, vol.~19, no.~1, pp. 44--57, Jan. 2002.

\bibitem{bioucas2012hyperspectral}
J.~M. Bioucas-Dias, A.~Plaza, N.~Dobigeon, M.~Parente, Q.~Du, P.~Gader, and
  J.~Chanussot, ``Hyperspectral unmixing overview: Geometrical, statistical,
  and sparse regression-based approaches,'' \emph{IEEE J. Sel. Topics Appl.
  Earth Observ. Remote Sens.}, vol.~5, no.~2, pp. 354--379, Apr. 2012.

\bibitem{ma2014signal}
W.-K. Ma, J.~M. Bioucas-Dias, T.-H. Chan, N.~Gillis, P.~Gader, A.~J. Plaza,
  A.~Ambikapathi, and C.-Y. Chi, ``A signal processing perspective on
  hyperspectral unmixing: Insights from remote sensing,'' \emph{IEEE Signal
  Process. Mag.}, vol.~31, no.~1, pp. 67--81, Jan. 2014.

\bibitem{heinz2001fully}
D.~C. Heinz \emph{et~al.}, ``Fully constrained least squares linear spectral
  mixture analysis method for material quantification in hyperspectral
  imagery,'' \emph{IEEE Trans. Geosci. Remote Sens.}, vol.~39, no.~3, pp.
  529--545, Mar. 2001.

\bibitem{boardman1993automating}
J.~W. Boardman, ``Automating spectral unmixing of aviris data using convex
  geometry concepts,'' in \emph{Proc. Summaries Annu. JPL Airborne Geosci.
  Workshop}, vol.~1, Nov. 1993, pp. 11--14.

\bibitem{winter1999n}
M.~E. Winter, ``N-findr: An algorithm for fast autonomous spectral end-member
  determination in hyperspectral data,'' in \emph{Proc. SPIE Conf. Imaging
  Spectrometry V}, vol. 3753, Oct. 1999, pp. 266--276.

\bibitem{nascimento2005vertex}
J.~M. Nascimento and J.~M. Bioucas-Dias, ``Vertex component analysis: A fast
  algorithm to unmix hyperspectral data,'' \emph{IEEE Trans. Geosci. Remote
  Sens.}, vol.~43, no.~4, pp. 898--910, Apr. 2005.

\bibitem{li2015minimum}
J.~Li, A.~Agathos, D.~Zaharie, J.~M. Bioucas-Dias, A.~Plaza, and X.~Li,
  ``Minimum volume simplex analysis: A fast algorithm for linear hyperspectral
  unmixing,'' \emph{IEEE Trans. Geosci. Remote Sens.}, vol.~53, no.~9, pp.
  5067--5082, Sep. 2015.

\bibitem{dobigeon2009joint}
N.~Dobigeon, S.~Moussaoui, M.~Coulon, J.-Y. Tourneret, and A.~O. Hero, ``Joint
  bayesian endmember extraction and linear unmixing for hyperspectral
  imagery,'' \emph{IEEE Trans. Signal Process.}, vol.~57, no.~11, pp.
  4355--4368, Nov. 2009.

\bibitem{eches2010bayesian}
O.~Eches, N.~Dobigeon, C.~Mailhes, and J.-Y. Tourneret, ``Bayesian estimation
  of linear mixtures using the normal compositional model. application to
  hyperspectral imagery,'' \emph{IEEE Trans. Image Process.}, vol.~19, no.~6,
  pp. 1403--1413, Jun. 2010.

\bibitem{nascimento2012hyperspectral}
J.~M. Nascimento and J.~M. Bioucas-Dias, ``Hyperspectral unmixing based on
  mixtures of dirichlet components,'' \emph{IEEE Trans. Geosci. Remote Sens.},
  vol.~50, no.~3, pp. 863--878, Mar. 2012.

\bibitem{qian2011hyperspectral}
Y.~Qian, S.~Jia, J.~Zhou, and A.~Robles-Kelly, ``Hyperspectral unmixing via
  ${L_{1/2}}$ sparsity-constrained nonnegative matrix factorization,''
  \emph{IEEE Trans. Geosci. Remote Sens.}, vol.~49, no.~11, pp. 4282--4297,
  Nov. 2011.

\bibitem{feng2018hyperspectral}
X.-R. Feng, H.-C. Li, J.~Li, Q.~Du, A.~Plaza, and W.~J. Emery, ``Hyperspectral
  unmixing using sparsity-constrained deep nonnegative matrix factorization
  with total variation,'' \emph{IEEE Trans. Geosci. Remote Sens.}, vol.~56,
  no.~10, pp. 6245--6257, Oct. 2018.

\bibitem{peng2021self}
J.~Peng, Y.~Zhou, W.~Sun, Q.~Du, and L.~Xia, ``Self-paced nonnegative matrix
  factorization for hyperspectral unmixing,'' \emph{IEEE Trans. Geosci. Remote
  Sens.}, vol.~59, no.~2, pp. 1501--1515, Feb. 2021.

\bibitem{zhu2014spectral}
F.~Zhu, Y.~Wang, B.~Fan, S.~Xiang, G.~Meng, and C.~Pan, ``Spectral unmixing via
  data-guided sparsity,'' \emph{IEEE Trans. Image Process.}, vol.~23, no.~12,
  pp. 5412--5427, Dec. 2014.

\bibitem{wang2017spatial}
X.~Wang, Y.~Zhong, L.~Zhang, and Y.~Xu, ``Spatial group sparsity regularized
  nonnegative matrix factorization for hyperspectral unmixing,'' \emph{IEEE
  Trans. Geosci. Remote Sens.}, vol.~55, no.~11, pp. 6287--6304, Nov. 2017.

\bibitem{huang2019spectral}
R.~Huang, X.~Li, and L.~Zhao, ``Spectral--spatial robust nonnegative matrix
  factorization for hyperspectral unmixing,'' \emph{IEEE Trans. Geosci. Remote
  Sens.}, vol.~57, no.~10, pp. 8235--8254, Oct. 2019.

\bibitem{iordache2011sparse}
M.-D. Iordache, J.~M. Bioucas-Dias, and A.~Plaza, ``Sparse unmixing of
  hyperspectral data,'' \emph{IEEE Trans. Geosci. Remote Sens.}, vol.~49,
  no.~6, pp. 2014--2039, Jun. 2011.

\bibitem{tang2015sparse}
W.~Tang, Z.~Shi, Y.~Wu, and C.~Zhang, ``Sparse unmixing of hyperspectral data
  using spectral a priori information,'' \emph{IEEE Trans. Geosci. Remote
  Sens.}, vol.~53, no.~2, pp. 770--783, Feb. 2015.

\bibitem{iordache2012total}
M.-D. Iordache, J.~M. Bioucas-Dias, and A.~Plaza, ``Total variation spatial
  regularization for sparse hyperspectral unmixing,'' \emph{IEEE Trans. Geosci.
  Remote Sens.}, vol.~50, no.~11, pp. 4484--4502, Nov. 2012.

\bibitem{iordache2014collaborative}
------, ``Collaborative sparse regression for hyperspectral unmixing,''
  \emph{IEEE Trans. Geosci. Remote Sens.}, vol.~52, no.~1, pp. 341--354, Jan.
  2014.

\bibitem{zheng2016reweighted}
C.~Y. Zheng, H.~Li, Q.~Wang, and C.~P. Chen, ``Reweighted sparse regression for
  hyperspectral unmixing,'' \emph{IEEE Trans. Geosci. Remote Sens.}, vol.~54,
  no.~1, pp. 479--488, Jan. 2016.

\bibitem{wang2017hyperspectral}
R.~Wang, H.-C. Li, A.~Pizurica, J.~Li, A.~Plaza, and W.~J. Emery,
  ``Hyperspectral unmixing using double reweighted sparse regression and total
  variation,'' \emph{IEEE Geosci. Remote Sens. Lett.}, vol.~14, no.~7, pp.
  1146--1150, Jul. 2017.

\bibitem{he2017total}
W.~He, H.~Zhang, and L.~Zhang, ``Total variation regularized reweighted sparse
  nonnegative matrix factorization for hyperspectral unmixing,'' \emph{IEEE
  Trans. Geosci. Remote Sens.}, vol.~55, no.~7, pp. 3909--3921, Jul. 2017.

\bibitem{zhang2018spectral}
S.~Zhang, J.~Li, H.-C. Li, C.~Deng, and A.~Plaza, ``Spectral--spatial weighted
  sparse regression for hyperspectral image unmixing,'' \emph{IEEE Trans.
  Geosci. Remote Sens.}, vol.~56, no.~6, pp. 3265--3276, Jun. 2018.

\bibitem{li2019local}
J.~Li, Y.~Li, R.~Song, S.~Mei, and Q.~Du, ``Local spectral similarity
  preserving regularized robust sparse hyperspectral unmixing,'' \emph{IEEE
  Trans. Geosci. Remote Sens.}, vol.~57, no.~10, pp. 7756--7769, Oct. 2019.

\bibitem{qi2020spectral}
L.~Qi, J.~Li, Y.~Wang, Y.~Huang, and X.~Gao, ``Spectral--spatial-weighted
  multiview collaborative sparse unmixing for hyperspectral images,''
  \emph{IEEE Trans. Geosci. Remote Sens.}, vol.~58, no.~12, pp. 8766--8779,
  Dec. 2020.

\bibitem{zheng2021sparse}
P.~Zheng, H.~Su, and Q.~Du, ``Sparse and low-rank constrained tensor
  factorization for hyperspectral image unmixing,'' \emph{IEEE J. Sel. Topics
  Appl. Earth Observ. Remote Sens.}, vol.~14, pp. 1754--1767, Jan. 2021.

\bibitem{licciardi2011pixel}
G.~A. Licciardi and F.~Del~Frate, ``Pixel unmixing in hyperspectral data by
  means of neural networks,'' \emph{IEEE Trans. Geosci. Remote Sens.}, vol.~49,
  no.~11, pp. 4163--4172, Nov. 2011.

\bibitem{palsson2018hyperspectral}
B.~Palsson, J.~Sigurdsson, J.~R. Sveinsson, and M.~O. Ulfarsson,
  ``Hyperspectral unmixing using a neural network autoencoder,'' \emph{IEEE
  Access}, vol.~6, pp. 25\,646--25\,656, May. 2018.

\bibitem{ozkan2019endnet}
S.~Ozkan, B.~Kaya, and G.~B. Akar, ``Endnet: Sparse autoencoder network for
  endmember extraction and hyperspectral unmixing,'' \emph{IEEE Trans. Geosci.
  Remote Sens.}, vol.~57, no.~1, pp. 482--496, Jan. 2019.

\bibitem{su2019daen}
Y.~Su, J.~Li, A.~Plaza, A.~Marinoni, P.~Gamba, and S.~Chakravortty, ``Daen:
  Deep autoencoder networks for hyperspectral unmixing,'' \emph{IEEE Trans.
  Geosci. Remote Sens.}, vol.~57, no.~7, pp. 4309--4321, Jul. 2019.

\bibitem{borsoi2019deep}
R.~A. Borsoi, T.~Imbiriba, and J.~C.~M. Bermudez, ``Deep generative endmember
  modeling: An application to unsupervised spectral unmixing,'' \emph{IEEE
  Trans. Comput. Imag.}, vol.~6, pp. 374--384, 2020.

\bibitem{qu2018udas}
Y.~Qu and H.~Qi, ``udas: An untied denoising autoencoder with sparsity for
  spectral unmixing,'' \emph{IEEE Trans. Geosci. Remote Sens.}, vol.~57, no.~3,
  pp. 1698--1712, Mar. 2019.

\bibitem{jin2021tanet}
Q.~Jin, Y.~Ma, X.~Mei, and J.~Ma, ``Tanet: An unsupervised two-stream
  autoencoder network for hyperspectral unmixing,'' \emph{IEEE Trans. Geosci.
  Remote Sens.}, vol. early access, 2021.

\bibitem{min2021jmnet}
A.~Min, Z.~Guo, H.~Li, and J.~Peng, ``Jmnet: Joint metric neural network for
  hyperspectral unmixing,'' \emph{IEEE Trans. Geosci. Remote Sens.}, vol. early
  access, 2021.

\bibitem{xiong2021snmf}
F.~Xiong, J.~Zhou, S.~Tao, J.~Lu, and Y.~Qian, ``Snmf-net: Learning a deep
  alternating neural network for hyperspectral unmixing,'' \emph{IEEE Trans.
  Geosci. Remote Sens.}, vol. early access, 2021.

\bibitem{mei2015equivalent}
S.~Mei, Q.~Du, and M.~He, ``Equivalent-sparse unmixing through spatial and
  spectral constrained endmember selection from an image-derived spectral
  library,'' \emph{IEEE J. Sel. Topics Appl. Earth Observ. Remote Sens.},
  vol.~8, no.~6, pp. 2665--2675, Jun. 2015.

\bibitem{mei2020improving}
S.~Mei, G.~Zhang, J.~Li, Y.~Zhang, and Q.~Du, ``Improving spectral-based
  endmember finding by exploring spatial context for hyperspectral unmixing,''
  \emph{IEEE J. Sel. Topics Appl. Earth Observ. Remote Sens.}, vol.~13, pp.
  3336--3349, Jun. 2020.

\bibitem{wang2021hyperspectral}
Q.~Wang, Q.~Li, and X.~Li, ``Hyperspectral image super-resolution using
  spectrum and feature context,'' \emph{IEEE Trans. Ind. Electron.}, vol.~68,
  no.~11, pp. 11\,276--11\,285, Nov. 2021.

\bibitem{zhang2021ssr}
X.~Zhang, W.~Huang, Q.~Wang, and X.~Li, ``Ssr-net: Spatial-spectral
  reconstruction network for hyperspectral and multispectral image fusion,''
  \emph{IEEE Trans. Geosci. Remote Sens.}, vol.~59, no.~7, pp. 5953--5965, Jul.
  2021.

\bibitem{palsson2019spectral}
B.~Palsson, J.~R. Sveinsson, and M.~O. Ulfarsson, ``Spectral-spatial
  hyperspectral unmixing using multitask learning,'' \emph{IEEE Access},
  vol.~7, pp. 148\,861--148\,872, 2019.

\bibitem{palsson2021convolutional}
B.~Palsson, M.~O. Ulfarsson, and J.~R. Sveinsson, ``Convolutional autoencoder
  for spectral--spatial hyperspectral unmixing,'' \emph{IEEE Trans. Geosci.
  Remote Sens.}, vol.~59, no.~1, pp. 535--549, Jan. 2021.

\bibitem{gao2021cycu}
L.~Gao, Z.~Han, D.~Hong, B.~Zhang, and J.~Chanussot, ``Cycu-net:
  Cycle-consistency unmixing network by learning cascaded autoencoders,''
  \emph{IEEE Trans. Geosci. Remote Sens.}, vol. early access, 2021.

\bibitem{hong2021endmember}
D.~Hong, L.~Gao, J.~Yao, N.~Yokoya, J.~Chanussot, U.~Heiden, and B.~Zhang,
  ``Endmember-guided unmixing network (egu-net): A general deep learning
  framework for self-supervised hyperspectral unmixing,'' \emph{IEEE Trans.
  Neural Netw. Learn. Syst.}, vol. early access, 2021.

\bibitem{rasti2021undip}
B.~Rasti, B.~Koirala, P.~Scheunders, and P.~Ghamisi, ``Undip: Hyperspectral
  unmixing using deep image prior,'' \emph{IEEE Trans. Geosci. Remote Sens.},
  vol. early access, 2021.

\bibitem{huang2020spatial}
Y.~Huang, J.~Li, L.~Qi, Y.~Wang, and X.~Gao, ``Spatial-spectral autoencoder
  networks for hyperspectral unmixing,'' in \emph{Proc. IEEE Int. Geosci.
  Remote Sens. Symp. (IGARSS'20)}, Sep. 2020, pp. 2396--2399.

\bibitem{zortea2009spatial}
M.~Zortea and A.~Plaza, ``Spatial preprocessing for endmember extraction,''
  \emph{IEEE Trans. Geosci. Remote Sens.}, vol.~47, no.~8, pp. 2679--2693, Aug.
  2009.

\bibitem{xu2018regional}
X.~Xu, J.~Li, C.~Wu, and A.~Plaza, ``Regional clustering-based spatial
  preprocessing for hyperspectral unmixing,'' \emph{Remote Sens. Environ.},
  vol. 204, pp. 333--346, Jan. 2018.

\bibitem{qi2019region}
L.~Qi, J.~Li, Y.~Wang, and X.~Gao, ``Region-based multiview sparse
  hyperspectral unmixing incorporating spectral library,'' \emph{IEEE Geosci.
  Remote Sens. Lett.}, vol.~16, no.~7, pp. 1140--1144, Jul. 2019.

\bibitem{li2021Superpixel}
H.~Li, R.~Feng, L.~Wang, Y.~Zhong, and L.~Zhang, ``Superpixel-based reweighted
  low-rank and total variation sparse unmixing for hyperspectral remote sensing
  imagery,'' \emph{IEEE Trans. Geosci. Remote Sens.}, vol.~59, no.~1, pp.
  629--647, Jan. 2021.

\bibitem{zhang2018hyperspectral}
X.~Zhang, Y.~Sun, J.~Zhang, P.~Wu, and L.~Jiao, ``Hyperspectral unmixing via
  deep convolutional neural networks,'' \emph{IEEE Geosci. Remote Sens. Lett.},
  vol.~15, no.~11, pp. 1755--1759, Nov. 2018.

\bibitem{qi2020deep}
L.~Qi, J.~Li, Y.~Wang, M.~Lei, and X.~Gao, ``Deep spectral convolution network
  for hyperspectral image unmixing with spectral library,'' \emph{Signal
  Process.}, vol. 176, p. 107672, Nov. 2020.

\bibitem{khajehrayeni2020hyperspectral}
F.~Khajehrayeni and H.~Ghassemian, ``Hyperspectral unmixing using deep
  convolutional autoencoders in a supervised scenario,'' \emph{IEEE J. Sel.
  Topics Appl. Earth Observ. Remote Sens.}, vol.~13, pp. 567--576, Feb. 2020.

\bibitem{eches2011enhancing}
O.~Eches, N.~Dobigeon, and J.-Y. Tourneret, ``Enhancing hyperspectral image
  unmixing with spatial correlations,'' \emph{IEEE Trans. Geosci. Remote
  Sens.}, vol.~49, no.~11, pp. 4239--4247, Nov. 2011.

\bibitem{giampouras2016simultaneously}
P.~V. Giampouras, K.~E. Themelis, A.~A. Rontogiannis, and K.~D. Koutroumbas,
  ``Simultaneously sparse and low-rank abundance matrix estimation for
  hyperspectral image unmixing,'' \emph{IEEE Trans. Geosci. Remote Sens.},
  vol.~54, no.~8, pp. 4775--4789, Aug. 2016.

\bibitem{li2018superpixel}
Z.~Li, J.~Chen, and S.~Rahardja, ``Superpixel construction for hyperspectral
  unmixing,'' in \emph{Proc. European Signal Process. Conf. (EUSIPCO)}, Sep.
  2018, pp. 647--651.

\bibitem{achanta2012slic}
R.~Achanta, A.~Shaji, K.~Smith, A.~Lucchi, P.~Fua, and S.~S{\"u}sstrunk, ``Slic
  superpixels compared to state-of-the-art superpixel methods,'' \emph{IEEE
  Trans. Pattern Anal. Mach. Intell.}, vol.~34, no.~11, pp. 2274--2282, Nov.
  2012.

\bibitem{clark2007usgs}
R.~N. Clark, G.~A. Swayze, R.~Wise, K.~E. Livo, T.~Hoefen, R.~F. Kokaly, and
  S.~J. Sutley, ``Usgs digital spectral library splib06a: Us geological survey,
  digital data series 231, 2007,'' \emph{URL http://speclab. cr. usgs.
  gov/spectral. lib06/ds231/index. html}, 2007.

\end{thebibliography}
\bibliographystyle{IEEEtran}

\end{document}